\documentclass[letterpaper,12pt]{article}
\pdfoutput=1 
\usepackage[bottom]{footmisc} 
\usepackage{amssymb}
\usepackage{amsthm}
\usepackage{newtxmath}
\usepackage{microtype} 
\usepackage[utf8]{inputenc}
\usepackage{newtxtext} 
\usepackage{setspace}
\usepackage{lineno,xcolor}
\usepackage{graphicx}
\usepackage{float}
\usepackage{printlen}
\usepackage{relsize}
\usepackage[hidelinks]{hyperref}
\usepackage{pdfpages}
\usepackage{makecell} 
\usepackage{placeins} 
\usepackage{rotating} 
\usepackage{tablefootnote} 

\usepackage[top=1.5in, bottom=1.5in, left=1in, right=1in]{geometry}
\usepackage[square,numbers,sort&compress]{natbib}
\usepackage[font={small}]{caption}
\usepackage{multirow}
\usepackage{url}

\usepackage{abstract} 

\usepackage{fancyhdr}
\fancyheadoffset{0cm}
\usepackage{titlesec} 
\usepackage{authblk} 
\usepackage{verbatim} 
\usepackage{sectsty}
\allsectionsfont{\boldmath}

\def\fwX{\href{http://fwx.pitt.edu}{\texttt{\textsc{fwXmachina}}}}
\def\TMVA{\href{http://root.cern/manual/tmva/}{\texttt{\textit{T}\textsc{MVA}}}}
\def\pT{p_\textrm{\scriptsize T}}
\def\ET{E_\textrm{\scriptsize T}}
\def\ETmiss{\ET^\textrm{miss}}
\def\ETmissj{E^\textrm{miss}_{\textrm{\scriptsize T},I}}
\def\SumET{\Sigma\ET}
\def\MET{\textrm{\small MET}}
\def\SET{\textrm{\small SET}}

\begin{document}

\thispagestyle{empty}
\title{
    \vspace{-15mm}
    \begin{flushright}{\large PITT-PACC-2212}\end{flushright}
    \vspace{10mm}
    \fontsize{18pt}{10pt}\selectfont\textbf{
    Nanosecond machine learning regression with deep\\ boosted decision trees in FPGA for high energy physics}
}

\author[a,b]{\fontsize{16pt}{10pt}\selectfont B.T.\ Carlson}
\author[b]{Q.\ Bayer}
\author[b]{\fontsize{16pt}{10pt}\selectfont T.M.~Hong\thanks{Corresponding author, tmhong@pitt.edu}}
\author[b]{S.T.\ Roche}
\affil[a]{\large Department of Physics, Westmont College }
\affil[b]{\large Department of Physics and Astronomy, University of Pittsburgh }
\date{\today}

\maketitle
\vspace{0mm}
\begin{abstract}
\noindent
    We present a novel application of the machine learning / artificial intelligence method called boosted decision trees to estimate physical quantities on field programmable gate arrays (FPGA).
    The software package \fwX\ features a new architecture called parallel decision paths that allows for deep decision trees with arbitrary number of input variables.
    It also features a new optimization scheme to use different numbers of bits for each input variable,
    which produces optimal physics results and ultraefficient FPGA resource utilization.
    Problems in high energy physics of proton collisions at the Large Hadron Collider (LHC) are considered.
    Estimation of missing transverse momentum ($\ETmiss$) at the first level trigger system at the High Luminosity LHC (HL-LHC) experiments,
    with a simplified detector modeled by Delphes,
    is used to benchmark and characterize the firmware performance.
    The firmware implementation with a maximum depth of up to $10$ using eight input variables of $16$-bit precision gives a latency value of $\mathcal{O}(10)\,\textrm{ns}$,
    independent of the clock speed,
    and $\mathcal{O}(0.1)\%$ of the available FPGA resources without using digital signal processors.
\end{abstract}
\vspace{18pt}
    Trigger concepts and systems (hardware and software).
\vfill

\newpage
\tableofcontents

\setlength\linenumbersep{15pt}
\renewcommand\linenumberfont{\normalfont\footnotesize\sffamily\color{gray}}
\modulolinenumbers[1]


\section{Introduction}

Fast,
accurate estimation of physical quantities from detector information is indispensable at 
high energy physics experiments,
especially ones with high data-taking rates.
At the Large Hadron Collider (LHC)~\cite{Evans:2008zzb},
for example,
bunches of protons collide at $40\,\textrm{MHz}$ that produce approximately $60\,\textrm{TB}$ of data each second.
At detectors such as those at the ATLAS~\cite{Aad:2008zzm} and CMS~\cite{Chatrchyan:2008aa} experiments,
an online trigger system saves a tiny fraction of the LHC collision events---%
often those deemed interesting---%
for follow-up analysis offline.
The data may include rare known physical processes sensitive to the effects from undiscovered laws of physics.

The trigger system must be capable of identifying such rare events to be saved offline while simultaneously rejecting the orders-of-magnitude larger background processes.
ATLAS and CMS experiments employ a two-level trigger system~\cite{Aad:2012xs,Aaboud:2016leb,Khachatryan:2016bia}.
The first level systems~\cite{Achenbach:2008zzb,Sirunyan:2020zal,CMS:2017gbu,Aad:2020hnu,ATLAS:2021tnq},
called level-0 or level-1 depending on the context,
have a latency requirement of a few microseconds per event to decide whether to save or reject the event \cite{CERN-LHCC-2017-020,Aad:1602235,CERN-LHCC-2020-004}.
Because of this strict requirement,
traditional algorithms often utilize simplified estimates,
relative to the offline algorithms,
at level-0\,/\,level-1.
Variables such as energy and momentum of a final state particle or invariant mass and angular correlations of a group of final state particles are typically considered.

For low latency implementation the firmware algorithms are used on custom electronic devices such as those that employ field programmable gate arrays (FPGA) and application-specific integrated circuits (ASIC).
A few thousandths of the incoming events pass the level-0\,/\,level-1 algorithms,
and pass on a rate of $\mathcal{O}(100)\,\textrm{kHz}$ of data to the software-based trigger system called high level trigger (HLT) or event filter (EF) depending on the context.
The HLT\,/\,EF uses a farm of CPUs to further evaluate events with more advanced algorithms within latency of a fraction of a millisecond.

Machine learning (ML) / artificial intelligence (AI) methods allow for improved estimates.
In particular,
regression models have been used for 
$\ETmiss$~\cite{CMS:2014xym},
tau energy estimation~\cite{ATLAS:2015boj},
electron and photon energy \cite{ATLAS:2018krz,ATLAS:2019qmc},
pileup mitigation~\cite{ATLAS:2019fxb},
and
pion energy calibration~\cite{ATLAS:2020efs}.

In the past few years,
progress in FPGA firmware design for signal-background classification have allowed for the use of more advanced algorithms using ML\,/\,AI 
at level-0\,/\,level-1
\cite{Duarte:2018ite,Jwa:2019zlh,Summers:2020xiy,DiGuglielmo:2020eqx,Iiyama:2020wap,Heintz:2020soy,Coelho:2020zfu,John:2020sak, Aarrestad:2021zos,Migliorini:2021fuj,Hong:2021snb,Elabd:2021lgo,Khoda:2022dwz},
typically relegated to the HLT\,/\,EF or offline analysis.
FPGA firmware design for regression estimates have been developed for experiments at level-0\,/\,level-1 \cite{Neuhaus:2014yma,Acosta:2018hjs,Aad:2021,Ospanov:2021EPJWC,Ospanov:2022fke}.

In this paper,
we present an alternative and novel FPGA firmware implementation of BDT algorithms that allows for deeper decision trees.
We expand on our previous BDT classification design \cite{Hong:2021snb} with $10\,\textrm{ns}$ latency and sub-percent-level resource usage and use it as a blueprint for regression.
We utilize the new design to perform regression to estimate physical quantities,
such as $\ETmiss$,
that may be of interest at the LHC.

The paper is organized as follows.
Section \ref{sec:ml_train} describes the regression problem and the machine learning training and setup.
Section \ref{sec:nano_opt} describes the nanosecond optimization,
which is the pre-processing step prior to the final firmware design.
Section \ref{sec:fw} describes the firmware design.
Section \ref{sec:results} presents the results.
Section \ref{sec:conclusion} concludes.

\section{ML training}
\label{sec:ml_train}

We consider the physics problem of the missing transverse momentum ($\ETmiss$). 
The $\ETmiss$ trigger serves as one of the primary means for identifying and saving high momentum particles that are invisible to the detector,
such as particles without strong or electromagnetic interactions.
Such particles include neutrinos as well as  those from hypothesized ``beyond'' the standard model (BSM) scenarios such as supersymmetry and dark matter (see, e.g., \cite{ATLAS:2019lng,ATLAS:2021kxv} and the references therein).

One particularly relevant example at ATLAS and CMS is the invisible decay of the Higgs bosons that are produced in vector boson fusion (VBF) during proton collisions \cite{Sirunyan:2018owy,ATLAS:2018bnv}.
In these cases the distribution of $\ETmiss$ decays relatively quickly above around $70\,\textrm{GeV}$ \cite{Buckley:2021gfw},
which makes it critical to maintain as low a trigger threshold as possible.
The experimental challenge is that the calorimeter noise level due to the high amount of ``pileup'' $\langle\mu\rangle$,
or simultaneous proton-proton collisions per bunch crossing,
drives the trigger thresholds to higher values to maintain a low event rate.
The pileup dependence of the $\ETmiss$ trigger has plagued ATLAS \cite{ATLAS:2020atr} and CMS \cite{CMS:2016ngn,CMS:2020cmk} during previous runs,
as the luminosity has rapidly increased and may continue to be problematic in future runs without new techniques to reduce pileup.
BDT regression is applied to estimate $\ETmiss$ defined below.

The calculation of $\ETmiss$ at the level-0\,/\,level-1 trigger system at the LHC is challenging because of the constraints imposed by the collider.
Examples of constraints include the strict timing between proton bunch crossings and the availability of detector information.
In $pp$ collisions,
the vector sum of momenta of the decay products in the plane normal to the beam should be zero.
In some collisions,
however,
momentum appears to not be conserved due to 
non-interacting decay products,
mismeasurement,
or a combination of both,
to produce a non-zero value of
\[
  \MET_I \equiv \ETmissj \equiv \Big|\sum_{i\,\in\,I} \vec{p}_\textrm{\scriptsize T,\,$i$}\ \Big|,
\]
where the sum is over the decay products.
For notational simplicity we also denote $\ETmiss$ as $\MET$.

\subsubsection*{Samples}

We created three samples \cite{samples},
two samples with non-zero distributions (items A1 and A2 below) and one sample with zero distribution (item B below) of 
true $\ETmiss$ at the generator level,
i.e.,
$\MET_\textrm{truth}$.
All three samples contain non-zero distributions of reconstructed $\MET$.

\begin{enumerate}
\item[A1.] Sample of Higgs bosons produced by VBF,
with the Higgs subsequently forced to decay to neutrinos via $H\rightarrow ZZ^\ast \rightarrow 4 \nu$,
producing non-zero $\MET_\textrm{truth}$.
The VBF process produces at least two highly energetic hadronic jets that are widely separated in polar angle with respect to the collision axis.
The Higgs decay to neutrinos ensures that the majority of the signal events would have a high value of $\MET$ at the generator level relative to the B sample below.
\item[A2.] Sample of leptonic $t\bar{t}$ decays with non-zero $\MET_\textrm{truth}$.
\item[B.\phantom{0}] Sample of the QCD multijet process with no $\MET_\textrm{truth}$.
\end{enumerate}

We note that samples A1 and B are merged into one training sample for the ML.
However,
for the evaluation of ROC curves where ``signal'' and ``background'' needs to be defined sample A1 is taken as the signal and sample B as the background.
An alternate choice of sample A2 as signal is considered to validate the training,
but we do not train with A2.

Each sample contains $100k$ events,
produced using MadGraph5$\_$aMC 2.9.5 \cite{Alwall:2014hca} for the matrix element calculation,
Pythia8 \cite{Sjostrand:2014zea} for the hadronization,
and Delphes 3.5.0~\cite{Ovyn:2009tx,deFavereau:2013fsa} for the detector simulation and object reconstruction.
The last step uses the ATLAS card with a mean pileup of $\langle\mu\rangle = 40$.

For the input variables described next,
the following objects are used.
Tracks from charged particles are reconstructed with a $\pT$ threshold of $100\,\textrm{MeV}$. 
Tracks have an efficiency applied as a function of their $\pT$ and location in $\eta$.
We use the default efficiency formulae from the ATLAS card with pileup in Delphes.
Neutral hadrons are summed into projective towers,
which are referred to as neutral hadron towers. 
Electrons and photons with a $\pT$ of at least $10\,\textrm{GeV}$ are reconstructed with an efficiency of $95\%$ within $|\eta|\,{<}\,1.5$ and $85\%$ within  $-2.5\,{<}\,\eta\,{<}\,-1.5$ and $1.5\,{<}\,\eta\,{<}\,2.5$.
The muons are the characterized similarly with the external $\eta$ boundary at $2.7$ instead of $2.5$.
Hadronic jets are reconstructed with the anti-$k_t$ algorithm \cite{Cacciari:2008gp} with a radius parameter $R=0.4$ and a minimum $\pT$ of $20\,\textrm{GeV}$.
For tracks,
the charged hadron subtraction method is used to remove pileup before jet reconstruction,
while the neutral component applies a residual correction to the reconstructed jet \cite{Cacciari:2007fd,Cacciari:2008gn}. 
The hadronic decays of tau leptons are reconstructed as jets.
A calorimeter with electromagnetic (EM) and hadronic (HAD) projective towers are formed, with 
fixed $\eta$-$\phi$ boundaries,
are distributed with a higher granularity in the central region $|\eta| < 2.6$ than in the forward region,
inspired by the granularity implemented in ATLAS~\cite{Aad:2010ai}.
Electrons and photons deposit all of their energy in the EM calorimeter,
while the energy from hadrons is split between the EM and HAD calorimeters.
Subsequently,
a detector energy resolution term is added separately for EM and HAD towers,
inspired by ref.\ \cite{Aharrouche:2006,Kulchitsky:2000gg}.
For the purpose of computing $\ETmiss$,
the energies from the EM and HAD towers are summed into a single tower.  
The plots also show the tower energy distributions for a typical signal and background event.

\subsubsection*{Variables}
We consider several algorithms for computing $\MET$,
which are modeled after those in use by ATLAS and CMS experiments during the Run 1 (2010--2012) and Run 2 (2015--2018) periods \cite{ATLAS:2018txj,CMS:2019ctu},
as input variables to the regression BDT.
We hypothesized that training a regression BDT on these inputs, similar to a regression constructed by CMS \cite{CMS:2014xym},
to target the generator-level $\MET$,
also called $\MET_\textrm{truth}$,
would yield a better approximate than any one input variable.

The input variables are four flavors of $\ETmiss$ 
($\MET_\textrm{reco}$,
$\MET_\textrm{towers}$,
$\MET_\textrm{tracks}$,
$\MET_\textrm{jets}$),
one flavor of $\SumET$ ($\SET_\textrm{jets}$),
and three energy densities
($\rho_\textrm{fwd-A}$,
$\rho_\textrm{barrel}$,
$\rho_\textrm{fwd-B}$).
The ``reconstructed'' $\ETmiss$ computed using objects with the $\pT$ thresholds described above
(electrons,
muons,
photons,
hadronic jets)
is $\MET_\textrm{reco}$.
The calorimeter tower-based $\ETmiss$ is $\MET_\textrm{towers}$.
The track- and neutral hadron tower-based $\ETmiss$ is $\MET_\textrm{tracks}$.
The jet-based $\ETmiss$ is $\MET_\textrm{jets}$;
this is sometimes referred to as $\textrm{MHT}$ in the literature.
The jet-based $\SumET$ is $\SET_\textrm{jets}$.
The energy density $\rho$ is computed from calorimeter towers in the forward regions and the barrel region of the calorimeter.

\begin{figure}[hbtp!]
    \centering
     \includegraphics[width=0.485\textwidth]{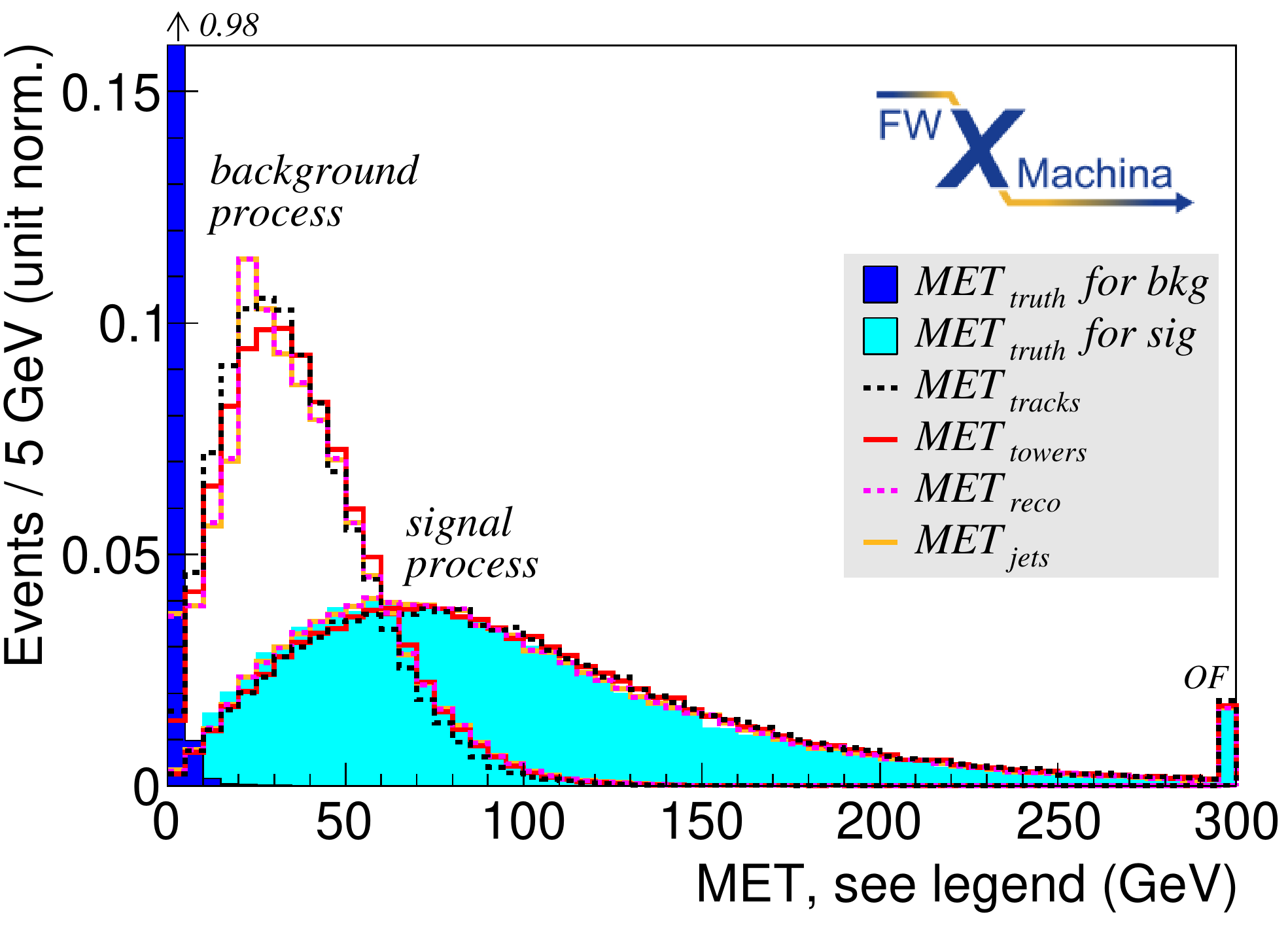}%
     \hspace{0.02\textwidth}%
     \includegraphics[width=0.485\textwidth]{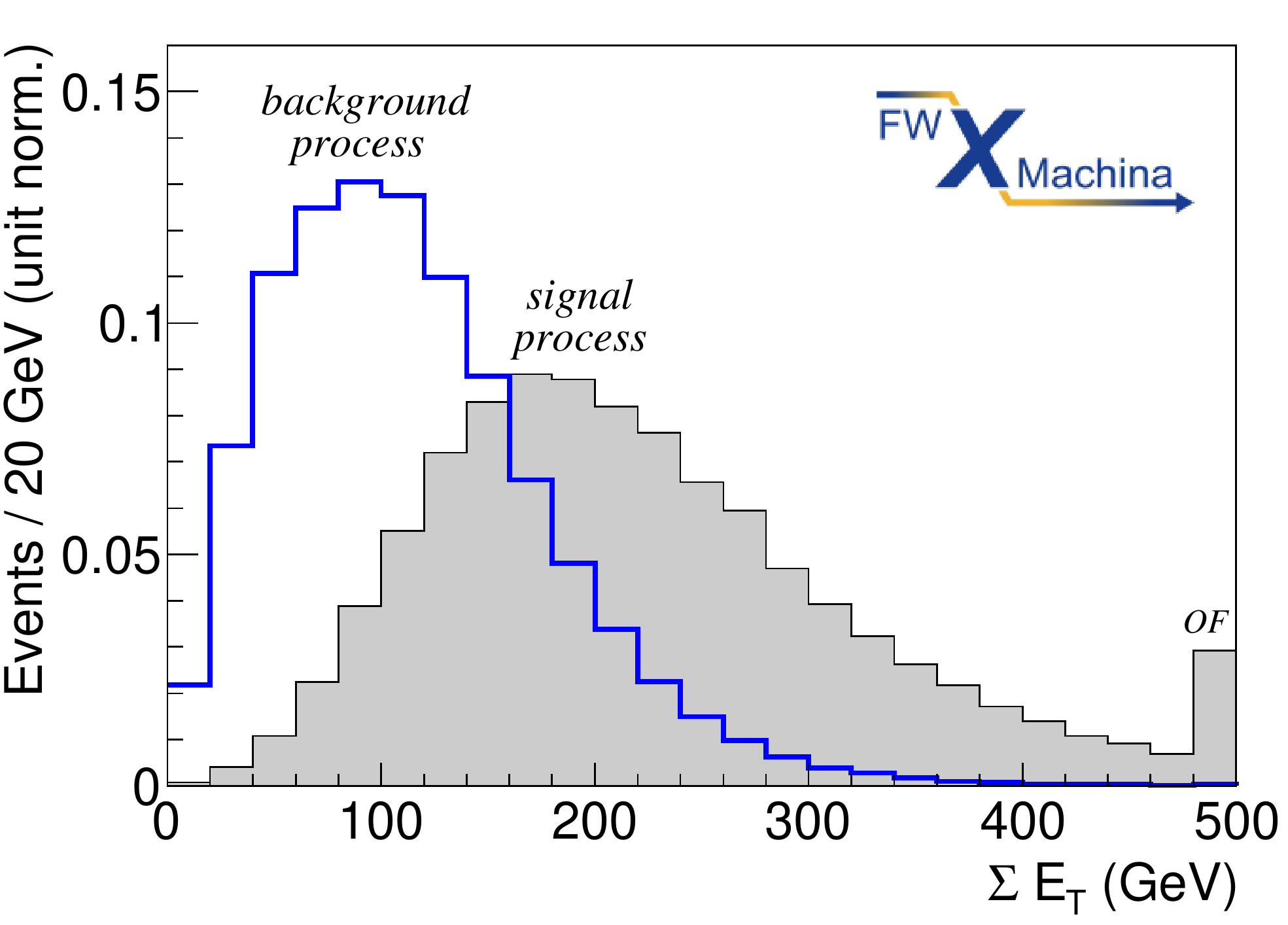}
    \caption{
        Input variable distributions.
        The reconstructed $\MET$ distributions (left) are given for background process for various estimations that all peak around $30\,\textrm{GeV}$ and the signal process that is broader that all peaks similarly at higher values.
        The truth $\MET$ distributions are also shown for the background process that peaks at $0$ and for the signal process that is similar to the reconstructed values.
        The reconstructed $\SumET$ distributions (right) are shown for the background process and the signal process.
    }
    \label{fig:inputs}
\end{figure}

Table~\ref{table:regression-variables} lists the input,
output,
and target variables for the regression BDT.
As mentioned previously,
the training is performed using one merged sample of sample A1 and sample B.
The input variable distributions are shown in figure \ref{fig:inputs}.
The plot on the left shows the various $\MET$ variables whereas the plot on the right shows the $\SET$ variable.

\begin{table}[htbp!]
\caption{
    List of variables for the $\ETmiss$ estimation.
    The regression takes eight input variables and optimizes to the target variable.
    The output is the result of the regression.
    \label{table:regression-variables}
}
\centering
{\small
\begin{tabular}{
    p{0.150\textwidth}
    p{0.150\textwidth}
    p{0.600\textwidth}
}
\hline
How used
& \makecell[tl]{Variable}
& Description
\\
\hline
Target       & $\MET_\textrm{truth}$     & $\ETmiss$ based on generator quantities due to, e.g., neutrinos \\
Input 1      & $\MET_\textrm{reco}$      & $\ETmiss$ based on reconstructed objects, e.g., $e$, $\mu$, $\gamma$, jets $j$ \\
~~~$''$~~~~2 & $\MET_\textrm{towers}$    & $\ETmiss$ based on calorimeter towers          \\
~~~$''$~~~~3 & $\MET_\textrm{tracks}$    & $\ETmiss$ by Delphes based on charged tracks and neutral hadron towers \\
~~~$''$~~~~4 & $\MET_\textrm{jets}$      & $\ETmiss$ based on reconstructed hadronic jets \\
~~~$''$~~~~5 & $\SET_\textrm{jets}$      & $\SumET$ of reconstructed hadronic jets        \\
~~~$''$~~~~6 & $\rho_\textrm{fwd-A}$     & Energy density for $-4.9 < \eta < -2.5$        \\
~~~$''$~~~~7 & $\rho_\textrm{barrel}$    & Energy density for $|\eta| < 2.5$              \\
~~~$''$~~~~8 & $\rho_\textrm{fwd-B}$     & Energy density for $2.5 < \eta < 4.9$          \\
Output       & $\mathcal{O}_\textrm{BDT}$& $\ETmiss$ estimation from the regression       \\
\hline
\end{tabular}
}
\end{table}

\subsubsection*{ML configuration}

The parameters of the regression BDT is determined by the \TMVA~\cite{Hocker:2007ht} library using the adaptive boosting (AdaBoost) method of regression variance separation.
It is configured with $40$ trees at a maximum depth of $5$ for the figures featuring ML results and the physics performance.

As is common with ML applications in the trigger system,
the training procedure is performed one time to determine the ML parameters.
The resulting setup is incorporated in the firmware design to evaluate the collision events in real-time.
The training step itself typically takes $\mathcal{O}(1)$ minute on a modern off-the-shelf laptop,
even for relatively deep networks.

The training step is performed on half of the events,
both VBF Higgs signal and multijet background samples together,
without pre-selection.
The other half is used to test the result of the training,
discussed in the next section.

\section{Nanosecond Optimization}
\label{sec:nano_opt}

Our BDT regression design is built on the framework of \fwX\ \cite{Hong:2021snb}.
The workflow is identical to the classification design with many components 
(\textsc{Tree Flattener},
\textsc{Forest Merger},
\textsc{Score Finder},
\textsc{Score Normalizer},
\textsc{Tree Remover},
and \textsc{Cut Eraser})
re-used for regression.
The new parallel decision path architecture is discussed in section \ref{sec:pdp}.

The treatment of the number of bits for variables is different than our previous design and is specific to the regression problem.
This is discussed later in \ref{sec:bits} and in appendix \ref{app:bits}.

\subsection{Parallel decision path architecture}
\label{sec:pdp}

The depth of the decision tree determines the granularity of the partitions of the input hyperspace:
the deeper the tree,
the finer the granularity.
We present a new non-iterative design of the deep decision tree with a complexity scaling with the depth and independent of the number of variables.
A comparison to our previous flattened design is given at the end of the section.

\begin{figure}[hbtp!]
{
\centering
\includegraphics[width=0.9\textwidth]{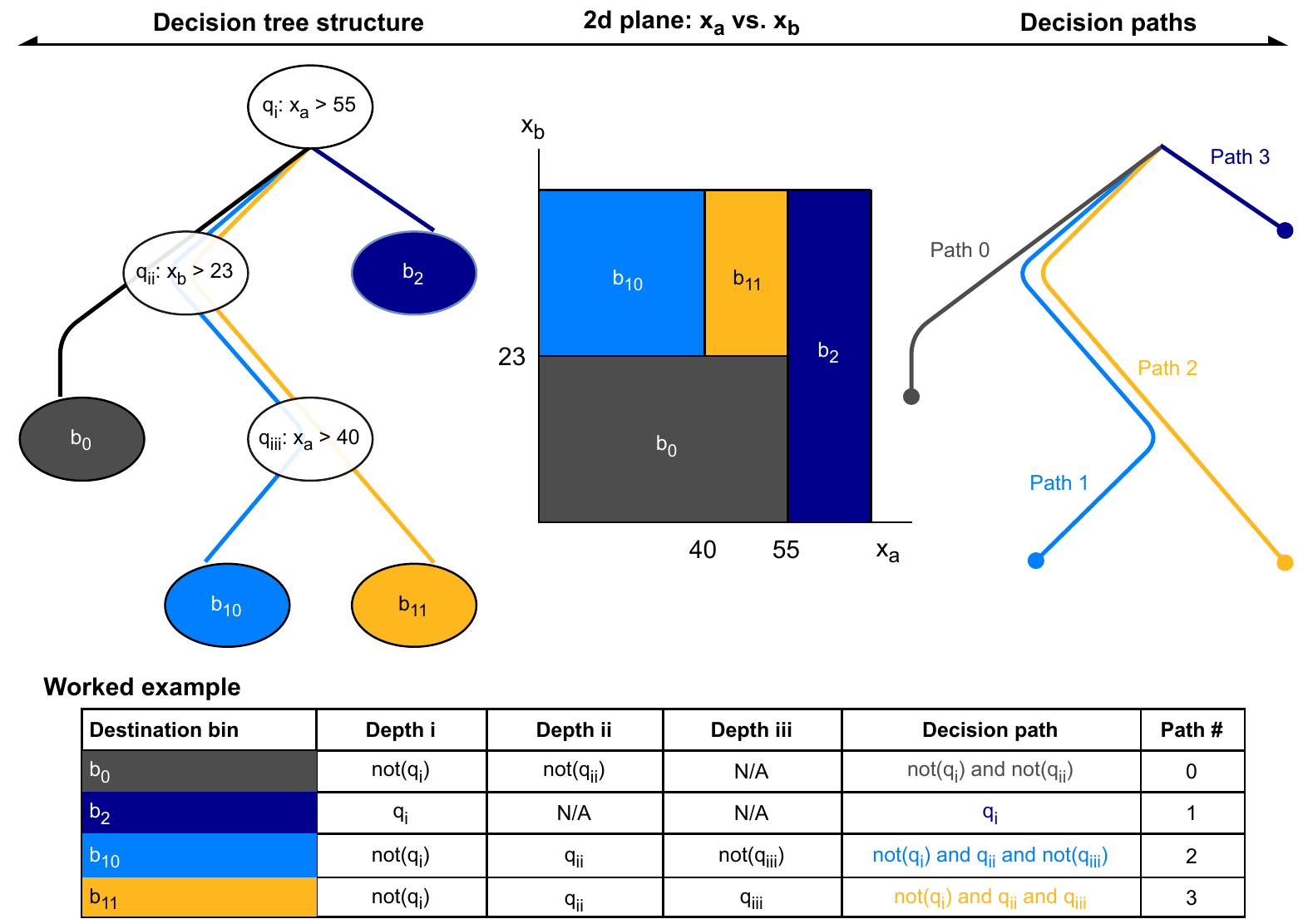}
\caption{
    Deep decision tree with parallel decision path (PDP) structure.
    An example is shown in the leftmost diagram for a decision tree using two variables ($x_a$, $x_b$) with a depth of $3$.
    The equivalent representation in the two-dimensional $x_a$ vs.\ $x_b$ space is given in the middle.
    The PDP perspective is given on the right.
    The table at the bottom lists the logical comparisons per PDP.
    \label{fig:dp}
    }
}
\end{figure}

Consider a decision tree of maximum depth $D$ with $N_\textrm{bin}=B$ terminal nodes, or bins.
By construction, $B$ is at most $2^D$.
The set of bins are labeled as $\{b_0, b_1, \ldots, b_{B-1}\}$.
Each terminal node is logically connected to the initial node by a set of comparisons that define the intermediate nodes.
For example,
a terminal node at depth $2$ corresponds to a set of $2$ comparisons.
More generally,
a terminal node $n$ at depth $d$ corresponds to a set of $d$ comparisons,
i.e.,
$Q_n = \{q^n_0,q^n_1,\ldots q^n_{d-1}\}$,
where $q^n_i$ is the result of the comparison at node $i$.
For a given terminal node $n$,
the logical $\textsc{and}$ of the associated set of comparisons is called a parallel decision path (PDP),
i.e.,
$\textrm{PDP}_n = {\displaystyle \cap}\,Q_n$.
Figure \ref{fig:dp} works out an example of one decision tree with two variables and maximum depth of three.

\begin{figure}[hbtp!]
{
\centering
\includegraphics[width=0.495\textwidth]{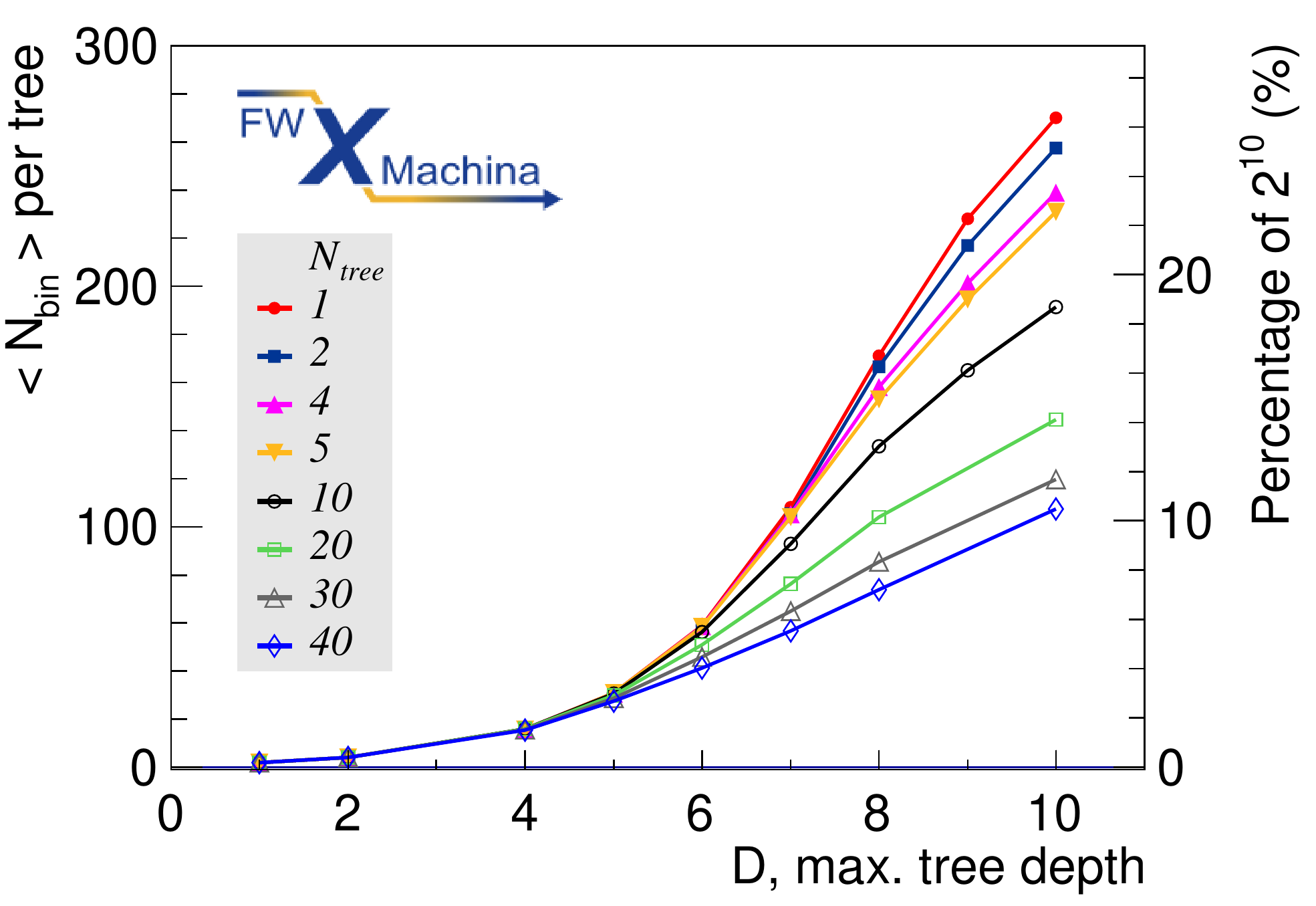}
\caption{
    Average number of bins per tree $\langle N_{bin}\rangle$ vs.\ maximum tree depth $D$.
    The right vertical axis shows the $\langle N_\textrm{bin}\rangle$ fraction with respect to the exponential scaling of $2^D$ to compare the points at $D=10$.
    \label{fig:bins_vs_d}
    }
}
\end{figure}

The advantage of the deep decision tree is that the set of decision paths can be evaluated simultaneously.
A fully populated tree with $2^D$ terminal notes has $2^D$ PDP and requires $D\,{\cdot}\,2^D$ comparisons.
We find that in our use cases,
deep trees are often not fully populated,
and thus scaling is often softer than $2^D$ (figure~\ref{fig:bins_vs_d}).
A comparison to the $2^D$ scaling for $D=10$ can be made with the vertical axis on the right.
We see that it ranges from $10$ to $25\%$ of a fully populated tree,
with density decreasing with $N_\textrm{tree}$.

\subsubsection*{Comparison to flattened trees architecture}

We compare the design in this paper with the non-iterative flattened architecture of ref.\ \cite{Hong:2021snb}.

One major limitation of the flattened architecture is that the number of bins $B$ scales with product of the number of cuts $c_v$ for each variable $v$ for $V$ total variables.
For even relatively shallow trees,
this quickly results in a prohibitive number of bins.
More formally,
one can get an idea of the scaling by considering the quantity $B = \prod_{v=1}^{V} c_v$ with some assumptions.
Each $c_v$ is bounded by $2^D$ since $\sum_{v=1}^{V}c_v=2^D-1$.
So if we suppose that each variable has the same number of cuts then,
we have $c_v\approx c$ where $c\equiv 2^D/V$.
In this scenario,
we have $B$ that scales exponentially with $D$ \underline{and} $V$.
In comparison,
PDP scales at worst exponentially with $D$ and independent of the number of variables $V$.
Moreover,
we saw a much softer scaling vs.\ $D$ in figure \ref{fig:bins_vs_d} in our examples.

\subsection{Variable number of bits}
\label{sec:bits}

Number of bits used per variable is an important aspect of resource optimization.
There are approaches that optimize prior to training \cite{Hong:2021snb} or during training \cite{Hawks:2021ruw}.
We take the former approach in optimally distributing the total numbers of bits to achieve the variable resolution at hand.
Some examples that are relevant to the problems at the LHC are discussed in appendix \ref{app:bits}.

In some applications of ML on FPGA,
it is beneficial if different variables can be represented with different bit integer precision.
For instance,
consider a sample BDT that uses two variables:
one with many cuts requiring exact precision,
and one with few cuts requiring less precision.
To minimize resource usage on the board,
it is useful if the variable with fewer cuts can be expressed with fewer bits.
In ref.~\cite{Hong:2021snb},
we introduced the ability to use a different precision for the input variables than the BDT output score.
Here,
we add support for different variable precision for each input variable.
The bit-optimized configuration gives the same timing results,
but with ultraefficient FPGA implementation,
with respect to the unoptimized configuration.

We present the following comparison to illustrate the difference.
In the unoptimized configuration we use a total of $144$ bits with $16$ bits for all input and output variables.
In the optimized configuration we use a total of $87$ bits with either $12$ or $5$ bits per variable depending on the variable.
In terms of physics results,
both configurations yield the same area under the ROC curve.
In terms of FPGA cost,
both configurations result in the same latency and interval measurements.
However,
the resource usage in look up tables (LUT) and flip flops (FF) are reduced by a factor of about $4$.
Table~\ref{table:bits} summarizes the results.

We expect similar significant reduction in resource usage when applied to the classification problems of ref.~\cite{Hong:2021snb}.

\begin{table}[p!]
\caption{\label{table:bits}
    Comparison of our benchmark configuration (details in table \ref{table:benchmark}) and the bit optimized configuration.
    The setup is given in the top;
    the physics performance in the middle;
    the FPGA cost in the bottom.
}
\centering
{\small
\begin{tabular}{
    p{0.42\textwidth}
    p{0.20\textwidth}
    p{0.15\textwidth}
    p{0.10\textwidth}
    }
\hline
Parameter   & Benchmark (table \ref{table:benchmark})
            & Bit optimized
            & Ratio \\
\hline
\multicolumn{3}{l}{Setup: no.\ of bits} \\
    \quad $\MET_k$, where $k\,{=}\,$reco, towers, tracks, jets
                                    & $16$ bits each & $12$ bits each & $1.3$ \\ 
    \quad $\SET_\textrm{jets}$      & $16$ bits & $12$ bits & $1.3$ \\
    \quad $\rho_k$, where $k\,{=}\,$fwd-A, barrel, fwd-B
                                    & $16$ bits each & $\phantom{0}5$ bits each & $3.2$ \\
    \quad $\mathcal{O}_\textrm{BDT}$& $16$ bits & $12$ bits & $1.3$ \\
    \quad All variables             & $144$ bits in total & $87$ bits in total & $1.7$ \\
\hline
\multicolumn{3}{l}{Physics performance}\\
    \quad Area under the ROC curve$^{\ref{fn:auc}}$ (AUC) & $0.90$ & $0.90$ & $1$ \\
\hline
\multicolumn{3}{l}{FPGA cost for 40 trees, 5 depth}\\
    \quad Latency                   & $6$ clock ticks & $6$ clock ticks  & $1$ \\
    \quad Interval                  & $1$ clock tick  & $1$ clock tick   & $1$ \\
    \quad Look up tables            & $1675$          & $374$            & $4.5$ \\
    \quad Flip flops                & $1460$          & $352$            & $4.1$ \\
    \quad Block RAM                 & $0$             & $0$              & Same \\ 
    \quad Ultra RAM                 & $0$             & $0$              & Same \\ 
    \quad Digital signal processors & $0$             & $0$              & Same \\ 
\hline
\end{tabular}
}
\end{table}

There are subtleties and technical challenges for the implementation.
The discussion can be found in appendix~\ref{app:bits}.
 
\section{Firmware design}
\label{sec:fw}

The firmware implementation of regression is built on \fwX\ \cite{Hong:2021snb}.
The only substantial difference of the existing pieces is in the \textsc{Score Processor} for the gradient boost (GradBoost).\footnote{%
    This component is intended to normalize the score provided by the BDT into a more useful form.
    The transformation for AdaBoost is still trivial, so this went unchanged.
    The GradBoost option now adds a supplied constant to the sum of the individual scores.
    The processor for GradBoost previously applied a piecewise approximation of the $\tanh$ function to the sum of the individual scores;
    however,
    this normalization method is no longer desirable.
    If no constant is provided when using GradBoost,
    then the constant is set to $0$ by default.
    }
A new \textsc{Engine} is introduced next for the parallel decision paths.
Firmware verification and validation is given in section \ref{sec:fw_valid}.

\subsection{\textsc{Deep Decision Tree Engine}}
\label{sec:ddte}

\begin{figure}[hbtp!]
{
\centering
\includegraphics[width=0.8\textwidth]{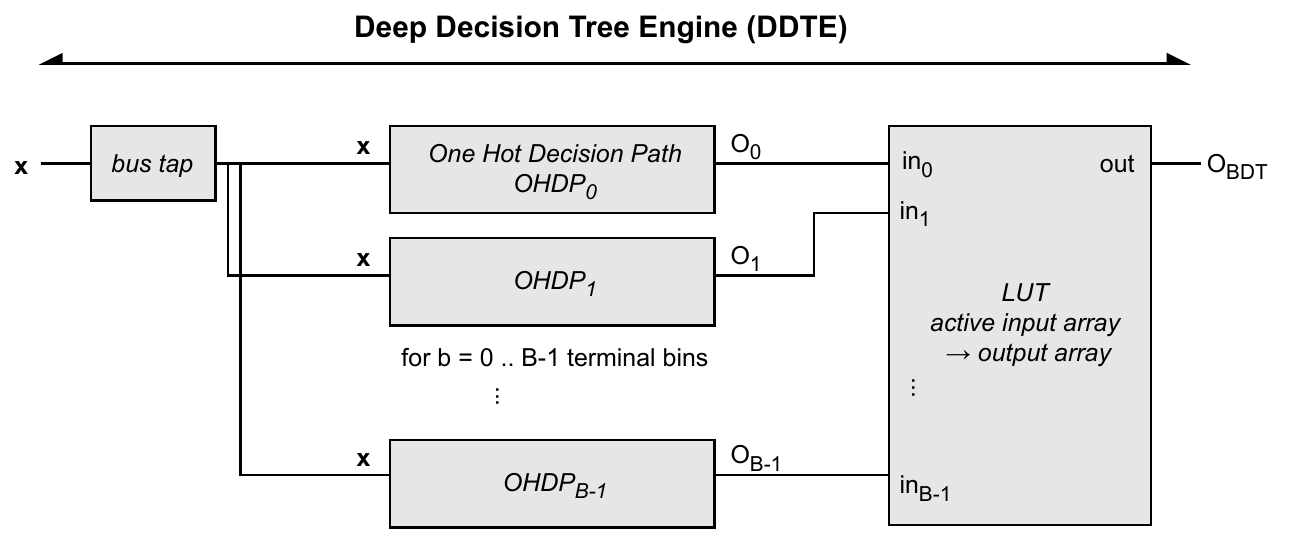}
\caption{
    The block diagram of the \textsc{Deep Decision Tree Engine} (DDTE).
    The input $\mathbf{x}$ is a vector of $V$ variables and the output $O_\textrm{BDT}$ is the regression estimate.
    For each $B$ bin (terminal node) of the decision tree,
    there is a corresponding \textsc{One Hot Decision Path} (OHDP).
    \label{fig:ddte}
    }
}
\end{figure}

Deep decision tree is implemented with a \textsc{Deep Decision Tree Engine} (DDTE);
see figure \ref{fig:ddte}.
DDTE makes $B$ duplicates of the set of input variables and feeds each one to the \textsc{One Hot Decision Path} (OHDP) to process each parallel decision path.
The OHDP results are collected by a look up table (LUT) that converts the one-hot vector into an regression estimate $O_\textrm{BDT}$.

\begin{figure}[hbtp!]
{
\centering
\includegraphics[width=0.8\textwidth]{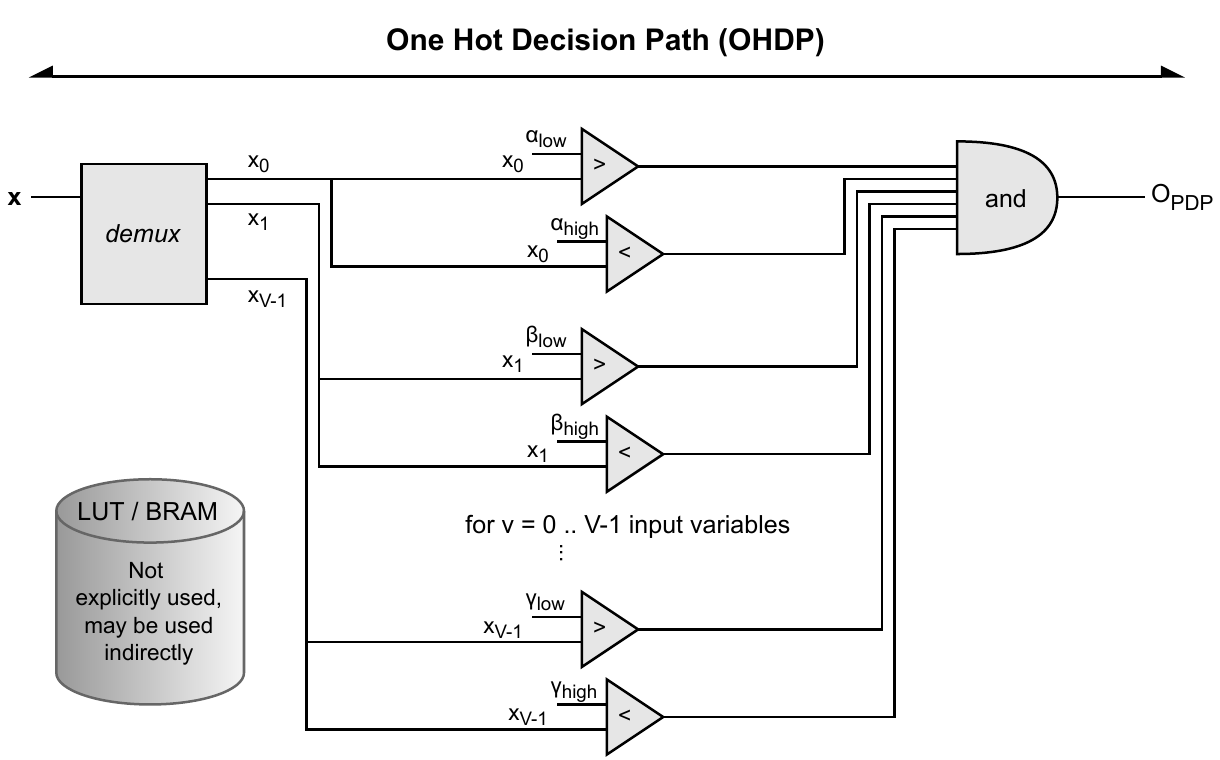}
\caption{
    The block diagram of the \textsc{One Hot Decision Path} (OHDP).
    Each variable $x_v$ is compared to pair of ``low'' and ``high'' constants $\alpha$ to check if it is within the range defined by the constants,
    i.e.,
    $\alpha_\textrm{low} < x_v < \alpha_\textrm{high}$.
    The collection of pairs of comparisons for each variable defines a particular parallel decision path (PDP).
    The output of OHDP is a boolean and is one-hot encoded.
    \label{fig:ohdp}
    }
}
\end{figure}

Each OHDP implements the parallel decision path logic with a pair of comparisons for each variable,
comparing to a user-configured upper bound and a lower bound;
see figure \ref{fig:ohdp}.
The output of each comparison feeds in to one $\textsf{and}$ operator to yield a binary result.

\subsection{Checks and comparisons}
\label{sec:fw_valid}

We report the results for the benchmark point of $40$ trees and a maximum depth of $5$ in this section.

\subsubsection*{Verification and validation}

For verification of the firmware against physical FPGA,
an RTL design is programmed onto the Xilinx Virtex UltraScale+ FPGA VCU118 Evaluation Kit and the Xilinx Artix-7 FPGA on Zynq-7020 System on Chip (SoC).
The Ultrascale+ is run on three clock speeds---%
$320\,\textrm{MHz}$,
$200\,\textrm{MHz}$,
and
$100\,\textrm{MHz}$---%
while the Artix-7 is run on $100\,\textrm{MHz}$.
In all scenarios,
a test vector is evaluated on the FPGA and the resulting output is compared to the output received from co-simulation.
No difference is observed.

Our firmware design produces latency values that are independent of the choice of clock speed,
as was also the case for ref.~\cite{Hong:2021snb}.
The same configuration is executed using the three clock speeds noted above on the Ultrascale+.
The results produced latency values of
$6$ clock ticks ($18.75\,\textrm{ns}$),
$4$ clock ticks ($20\,\textrm{ns}$),
and
$2$ clock ticks ($20\,\textrm{ns}$),
respectively.
They all produce the same latency values of about $20\,\textrm{ns}$ within the resolution of the clock period.
The interval remains at one clock tick for all clock speeds.

The latency for the Artix-7 was about 4-fold higher,
in absolute terms using the same clock speed,
than for the Ultrascale+ as was seen previously for our flattened tree architecture \cite{Hong:2021snb}.
Resource cost was generally higher as well on the Artix-7 with over 6-fold the resource usage compared to the Ultrascale+.
As is seen for the Ultrascale+ results,
the interval is one clock tick and no DSP is used.

Validation of the firmware against software simulation is done with $100\,000$ input test vectors for around $200$ different BDT configurations.
Inputs are processed through software simulation as well as co-simulation.
Other than rounding discrepancies,
no difference is observed for the BDT output.

\subsubsection*{Comparison of estimated and actual FPGA cost} 

We compare the actual FPGA resource utilization and timing results to the estimated values we obtain from C synthesis.
The actual values can be measured in Xilinx Vivado after the RTL design is generated and uploaded to the FPGA.
Timing is measured using the ILA,
while the resource utilization is directly reported by the software. 
We refer to figure 24 in ref.~\cite{Hong:2021snb} for the illustration of the definitions.

For the timing,
we see no difference between the estimated and actual values in all of the configurations that we considered in this paper.

\begin{figure}[hbtp!]
    \centering
    \includegraphics[width=0.495\textwidth]{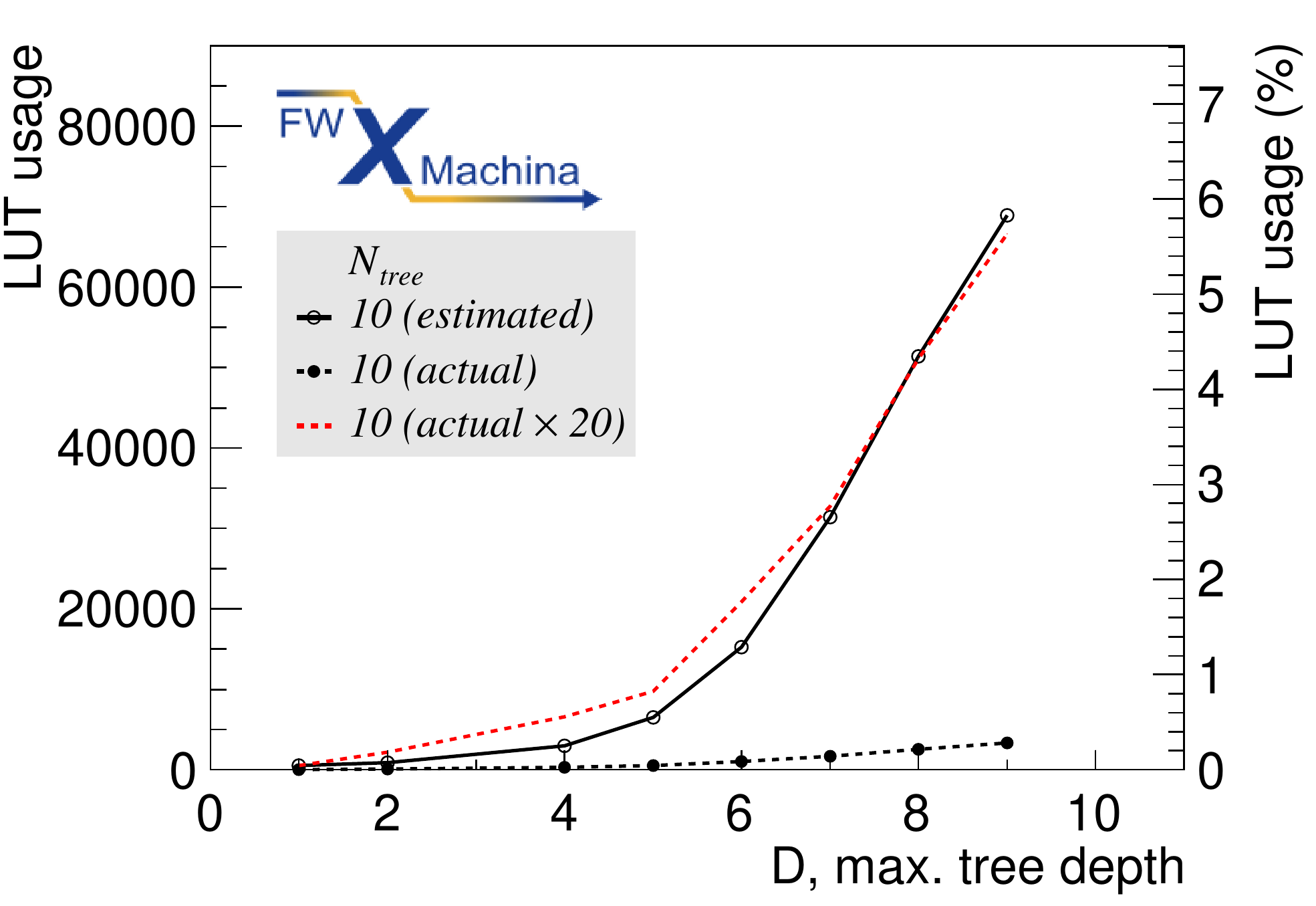}\hspace{0.01\textwidth}%
    \includegraphics[width=0.495\textwidth]{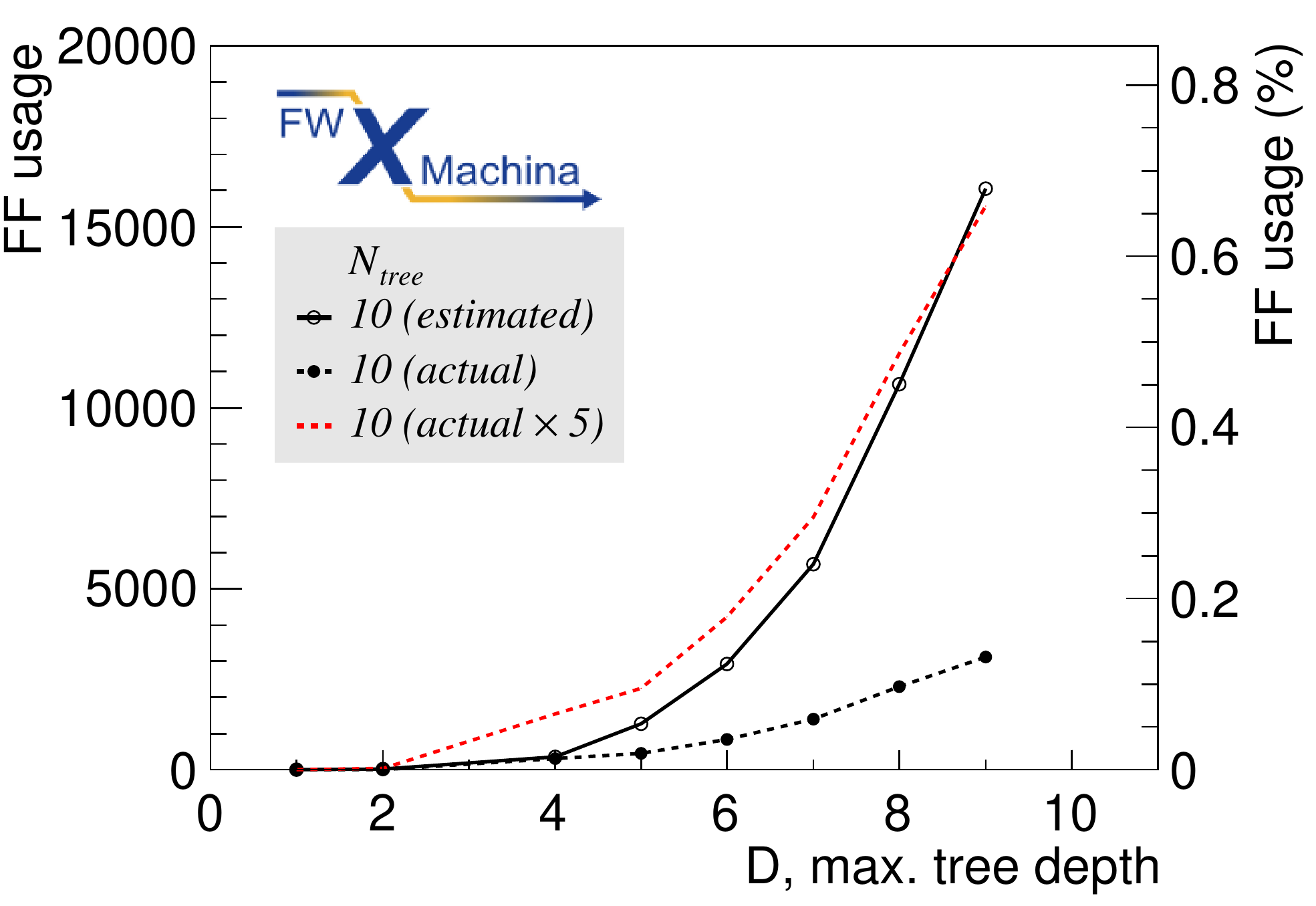}
    \\
    \includegraphics[width=0.495\textwidth]{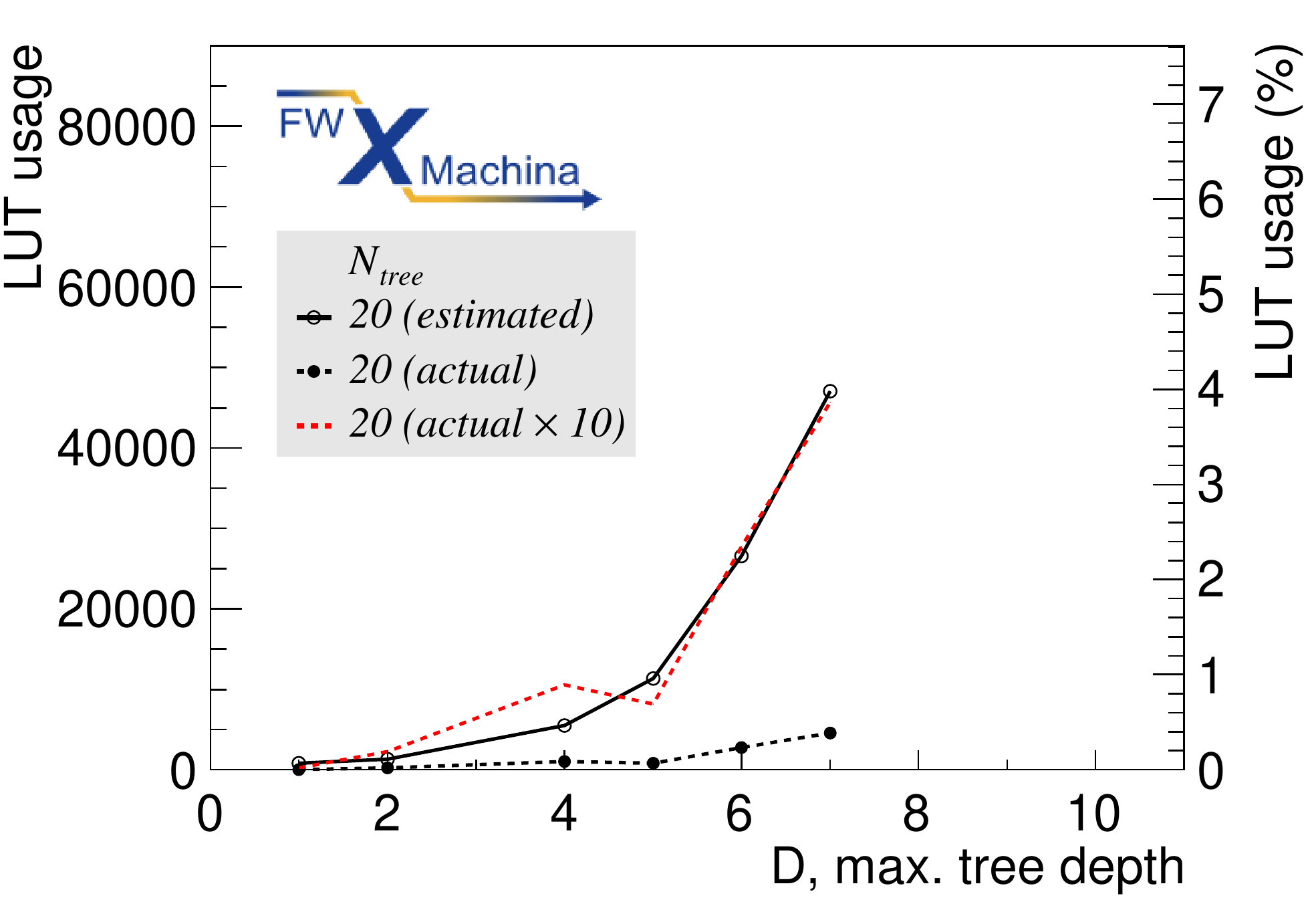}\hspace{0.01\textwidth}%
    \includegraphics[width=0.495\textwidth]{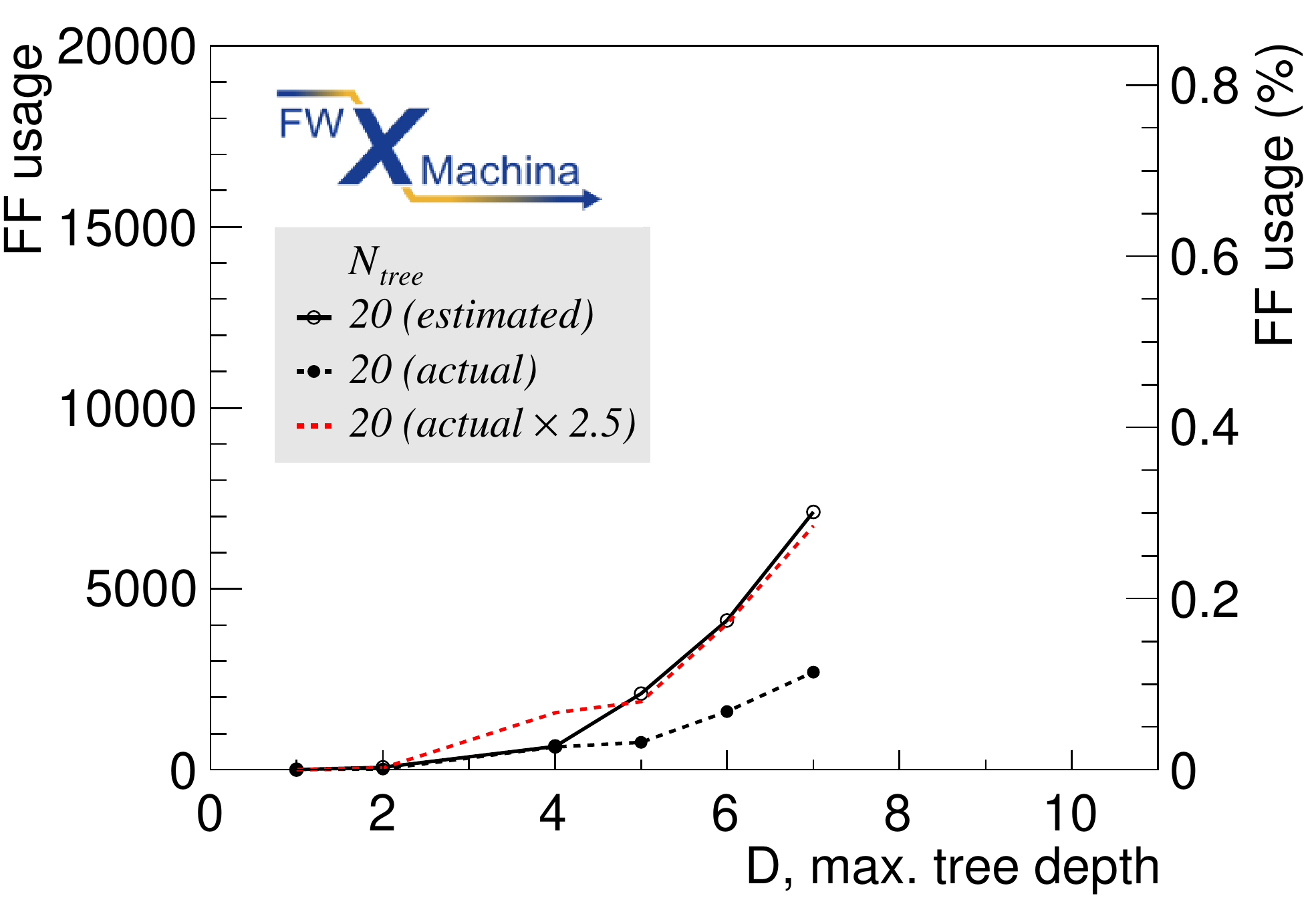}
    \caption{
    Comparison of estimated usage and actual usage
    for LUT (left column) and FF (right column)
    for $N_\textrm{tree}=10$ (top row).
    and $N_\textrm{tree}=20$ (bottom row).
    Estimated values are obtained with HLS C synthesis and the actual values are obtained by RTL synthesis and implementation.
    }
    \label{fig:fpga_vs_csynthesis}
\end{figure}

For the resources,
the resource utilization on the FPGA was consistently lower than C synthesis values. 
Figure~\ref{fig:fpga_vs_csynthesis} shows the LUT and the FF comparisons for the configuration using $N_\textrm{tree}=10$ and otherwise the same setup as in table~\ref{table:benchmark}.
It is notable that the actual values are lower than the estimated  values by a factor of about $20$ for the LUT and about $5$ for the FF.
We repeated the exercise for $N_\textrm{tree}=20$ and the scale factors are about $10$ for the LUT and about $2.5$ for the FF.
The actual values are scaled up by the stated factor,
which shows that the scaling vs.\ tree depth follows the trend presented by the estimated curve.
For this reason we report the actual FPGA resource utilization,
rather than the estimated version,
in section \ref{sec:results}.

\section{Results}
\label{sec:results}

We present the physics performance followed by the FPGA cost (timing and resource utilization) for the $\ETmiss$ problem.

\subsubsection*{Physics performance}

Physics performance is evaluated with ROC curves,
turn-on curves,
and $\MET$ resolution.

\begin{figure}[hbtp!]
    \centering
    \includegraphics[width=0.495\textwidth]{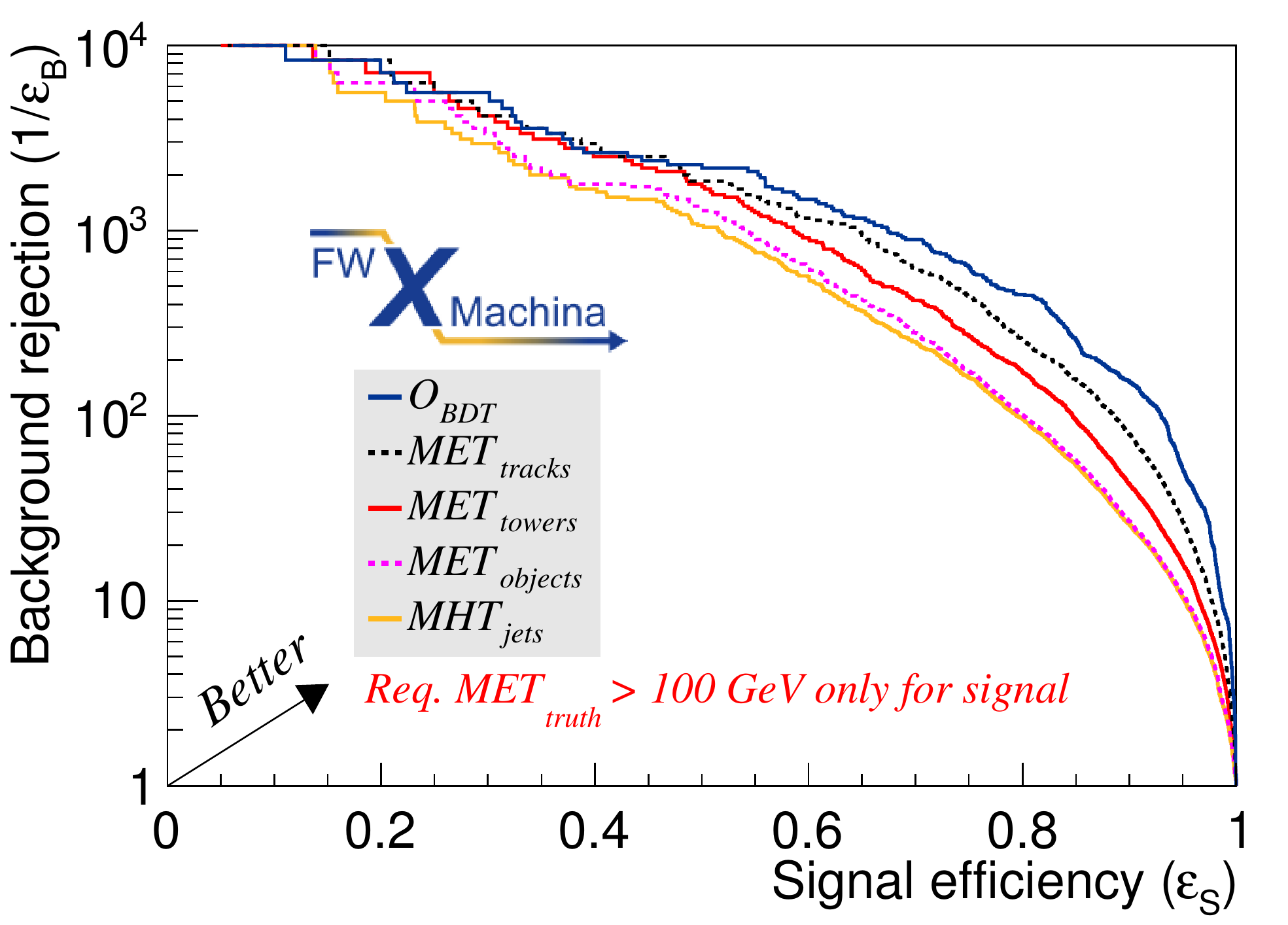}\hspace{0.01\textwidth}%
    \includegraphics[width=0.495\textwidth]{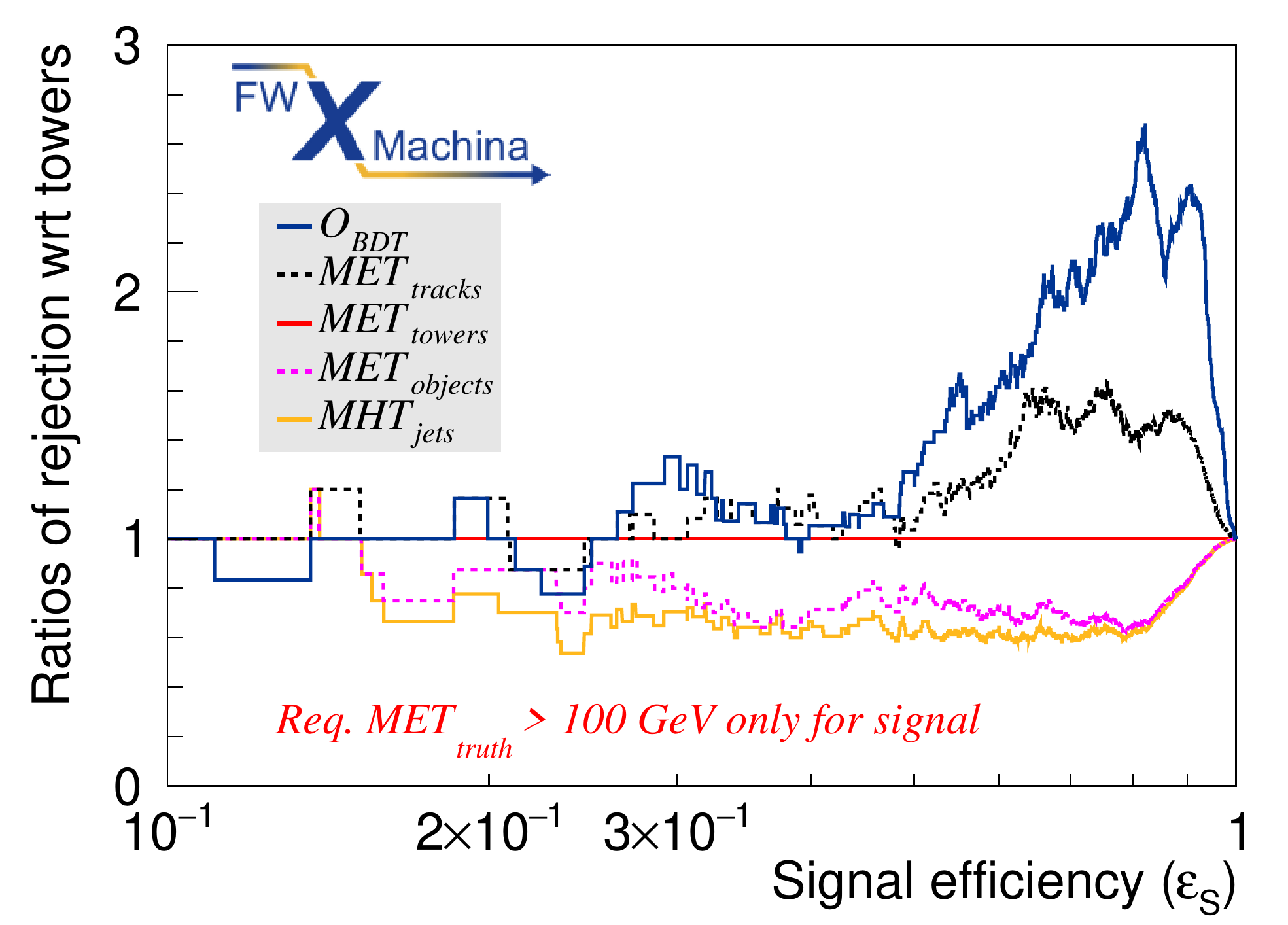}
    \caption{
        ROC curves of background rejection factor vs.\ signal acceptance (left) and
        the ratio of rejection factors with respect to $\MET_\textrm{towers}$ vs.\ signal acceptance (right).
        The background and signal are trained and evaluated using the QCD multijet and the VBF Higgs to $4\nu$ sample.
        The scan is done over the full $\MET$ range from $0$ to $1.5\,\textrm{TeV}$.
        The order of the legend follows the curves.
        A subset of events for which the signal sample has the pre-selection  $\MET_\textrm{truth}>100\,\textrm{GeV}$ applied is shown.
        The BDT values are using 16-bits.
    }
    \label{fig:roc}
\end{figure}

ROC curves are shown in left plot of figure~\ref{fig:roc},
showing the background rejection factor vs.\ signal acceptance.
The former is defined as the inverse of the background efficiency $1/\varepsilon_B$,
where $\varepsilon_B$ is the false positive rate (FPR) or type I error ($\alpha$).
The latter is defined as the signal efficiency $\varepsilon_S$,
also called the true positive rate (TPR).
The efficiencies for category $c=S,B$ are defined as the ratio of the number of events category $c$ passing the $\MET$ threshold with respect to the total number of events of category $c$,
i.e.,
$\varepsilon_c\equiv N_c^\textrm{pass}/N_c^\textrm{total}$.
A scan of the $\MET$ thresholds correspond to a point in the $(\varepsilon_S,1/\varepsilon_B)$ plane;
the collection of points define the ROC curve shown in the figure.

The right plot of figure \ref{fig:roc} shows the ratio of background rejection factors with respect to $\MET_\textrm{towers}$.
Plots in figure \ref{fig:roc} impose a selection of $\MET_\textrm{truth}\,{>}\,100\,\textrm{GeV}$ for the signal events to better illustrate the impact of the BDT for larger values of background rejection,
closer to a more realistic experimental threshold. 

\begin{figure}[hbtp!]
    \centering
    \includegraphics[width=0.495\textwidth]{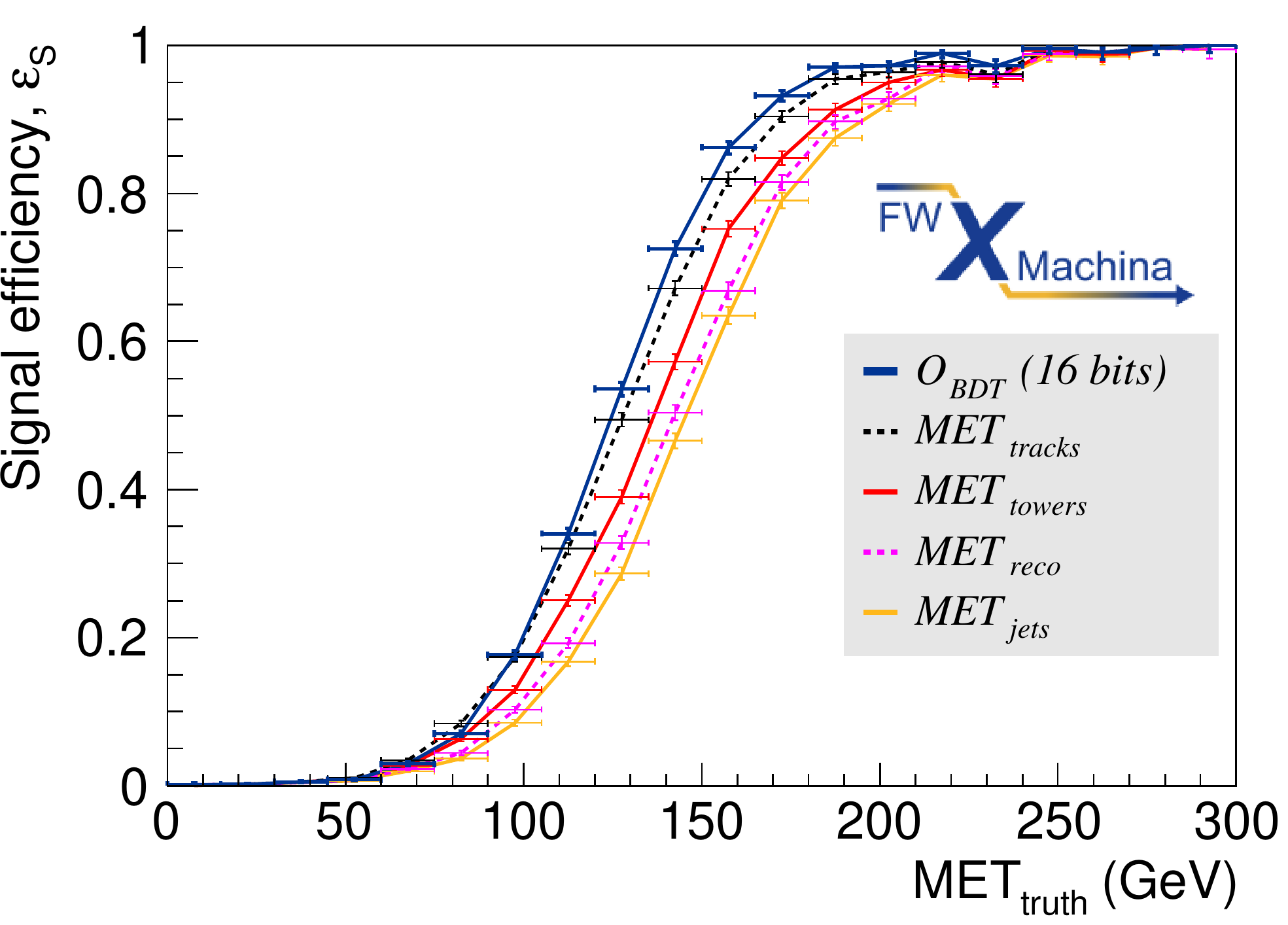}\hspace{0.005\textwidth}%
    \includegraphics[width=0.495\textwidth]{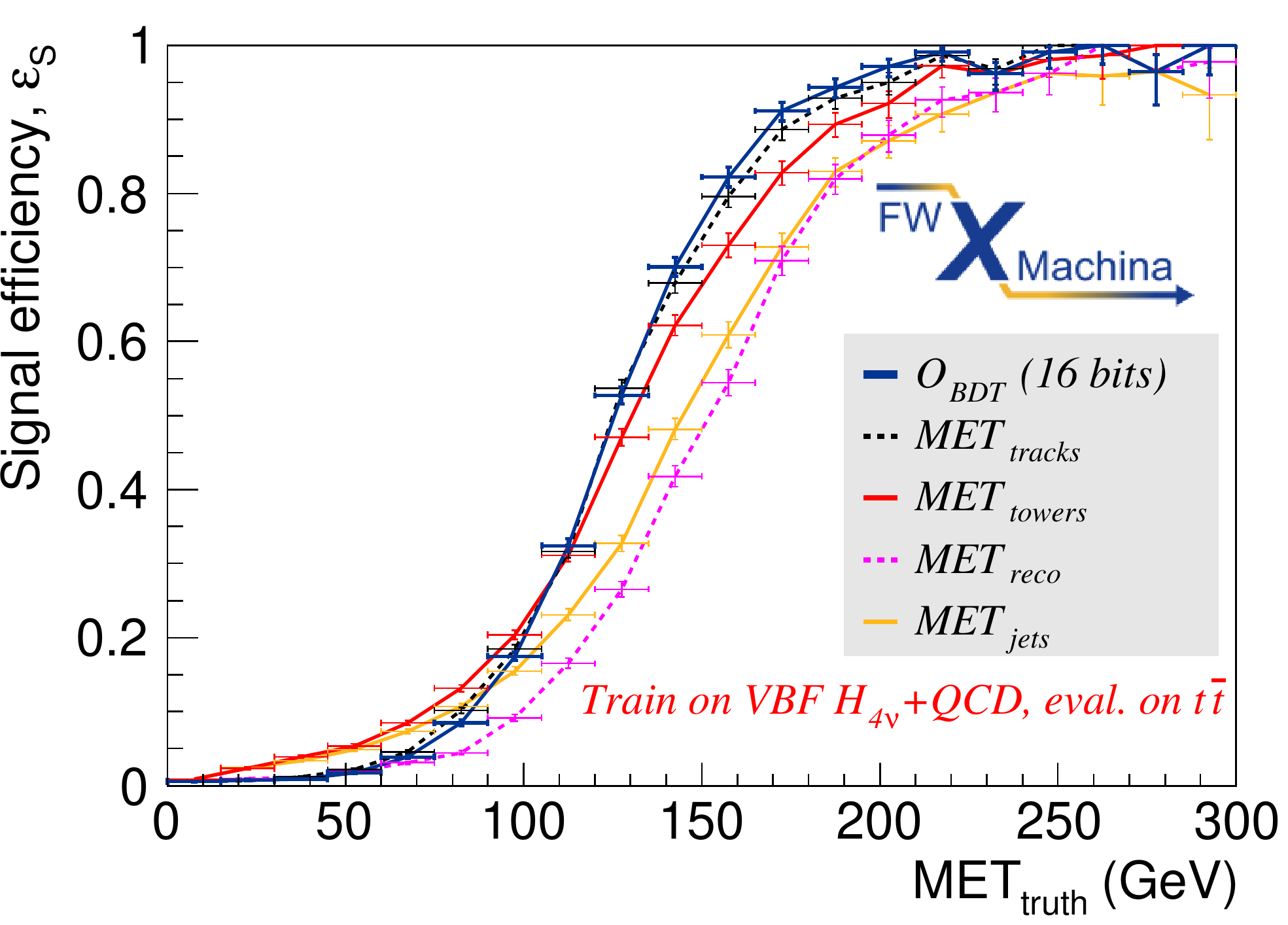}
    \caption{
    Efficiency turn-on curves for the VBF Higgs sample (left) and $t\bar{t}$ samples (right) as signal on the vertical axis and QCD multijet for background on the horizontal axis.
    The signal efficiency of the BDT is improved for both VBF Higgs and $t\bar{t}$ samples as signal,
    demonstrating that the regression is robust for a wide range of topologies.
    The threshold corresponding to the operating point of background efficiency of $\varepsilon_B=10^{-3}$ is chosen.
    For each histogram a line is drawn between the data points as a visual guide.
    }
    \label{fig:turnon}
\end{figure}

For an operating point on the ROC curve---%
say, at a background efficiency of $\varepsilon_B=10^{-3}$---%
the signal efficiency can be read off of figure \ref{fig:roc}.
For a signal efficiency value of,
say $85\%$,
the background rejection is approximately $150\%$ lower than the same efficiency computed using $\MET$ formed only with calorimeter towers. 
More information can be obtained at that operating point by scanning the $\MET_\textrm{truth}$ threshold given an algorithm on the signal process.
This produces the so-called turn-on curve in the left plot of figure \ref{fig:turnon}.

The turn-on curve is evaluated by identifying a selection of the $\MET$ variable (e.g., $\mathcal{O}_\textrm{BDT} > 75\,\textrm{GeV}$),
and for each bin of $\MET_\textrm{truth}$ evaluate the fraction of events that satisfy the selection. 
The BDT outperforms other curves by reaching full efficiency the quickest,
i.e.,
the turn-on curve is ``sharper.''

In order to verify the performance of an alternate sample with non-zero $\MET_\textrm{truth}$ with a larger jet multiplicity,
including jets in the central region unlike for VBF processes,
we consider leptonic $t\bar{t}$ decays in the right plot of figure \ref{fig:turnon}.
The same BDT trained on the merged sample of sample A1 and B is used to evaluate sample A2 for $t\bar{t}$.
We see that that turn-on curve for the BDT is sharper than the input $\MET$ variables.

\begin{figure}[hbtp!]
    \centering
    \includegraphics[width=0.495\textwidth]{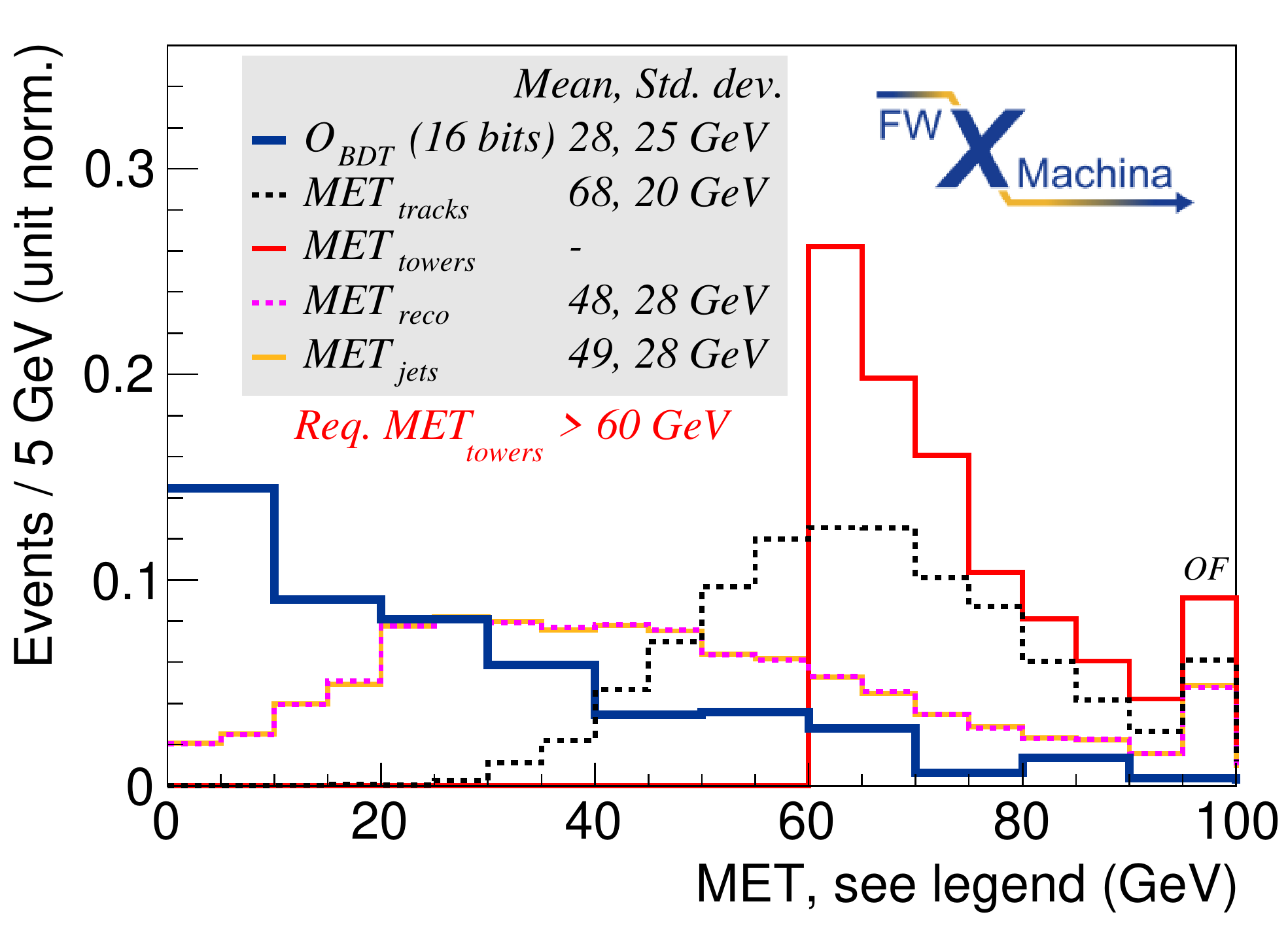}\hspace{0.01\textwidth}%
    \includegraphics[width=0.495\textwidth]{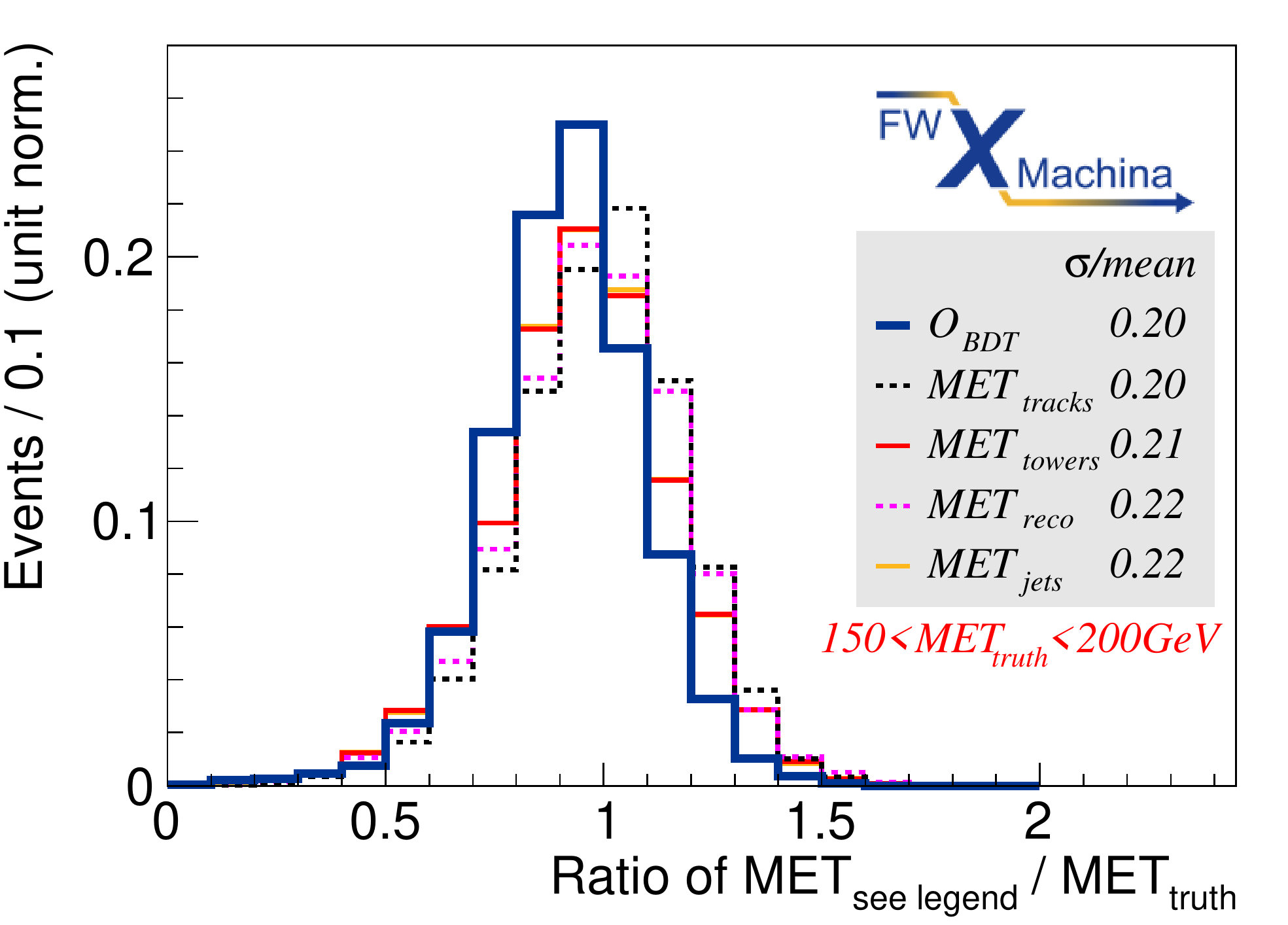}
    \caption{
        $\ETmiss$ distributions for background events (left) and $\ETmiss$ resolution for signal events (right).
        Background events (left) with $\MET_\textrm{towers}\,{>}\,60\,\textrm{GeV}$ shows how the higher $\MET$ values get remapped.
        Signal events (right) with $100<\MET_\textrm{truth}<200\,\textrm{GeV}$ shows the resolution with $\MET_\textrm{truth}$ values of interest.
        The input variable distributions are shown using floating point values.
        The output estimate $\mathcal{O}_\textrm{BDT}$ distributions are shown using 16-bit as is done in firmware.
    }
    \label{fig:met}
\end{figure}

Finally,
we consider $\MET$ resolution.
If the algorithm rejects more background events while retaining a similar amount of signal events,
the BDT regression is worth pursuing.
We will see that this is the case.
The $\MET$ distribution for events without true $\ETmiss$,
i.e., background events,
is shown in the left plot of figure~\ref{fig:met}.
As the training sample includes background events without true $\ETmiss$,
these events tend to be reconstructed with low values of $\MET$ by the regression model,
as expected.
To highlight the improved rejection, 
subset of events with non-zero reconstructed $\MET$,
taken to be $\MET_\textrm{towers}>60\,\textrm{GeV}$ are selected.
For this subset of events, 
the BDT estimate outperforms the input $\MET$ variables for accurately estimating the null $\MET$ value.

The $\MET$ resolution for events with true $\ETmiss$,
i.e.,
signal events,
is shown in the right plot of figure~\ref{fig:met}.
For the subset of events with non-zero reconstructed $\MET$ in the range at which a $\MET$ cut might be placed in a trigger system,
taken here to be $150<\MET_\textrm{truth}<200\,\textrm{GeV}$,
the BDT estimate performs comparably to the input $\MET$ variables.

\begin{figure}[hbtp!]
    \centering
    \includegraphics[width=0.495\textwidth]{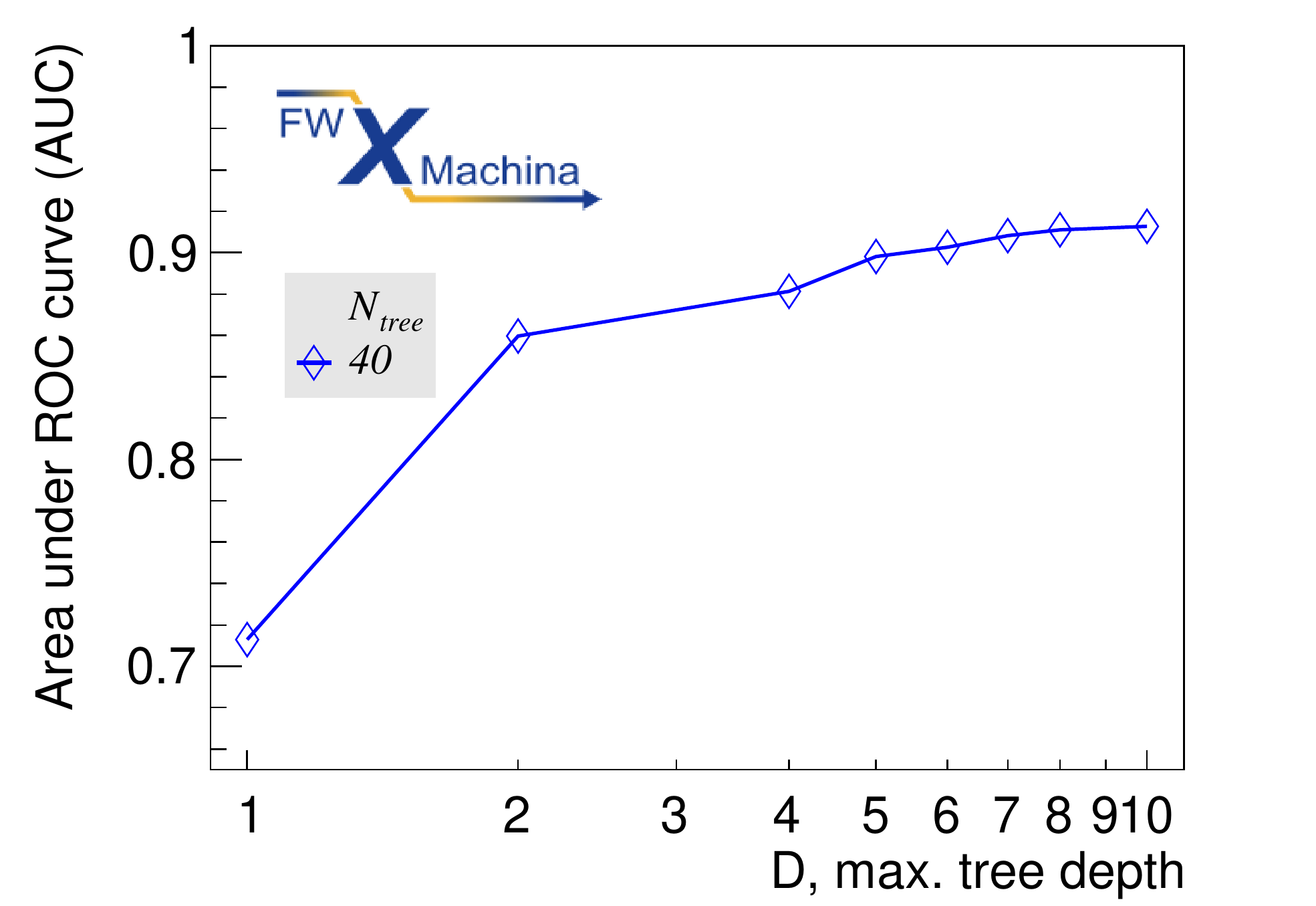}\hspace{0.005\textwidth}%
    \includegraphics[width=0.495\textwidth]{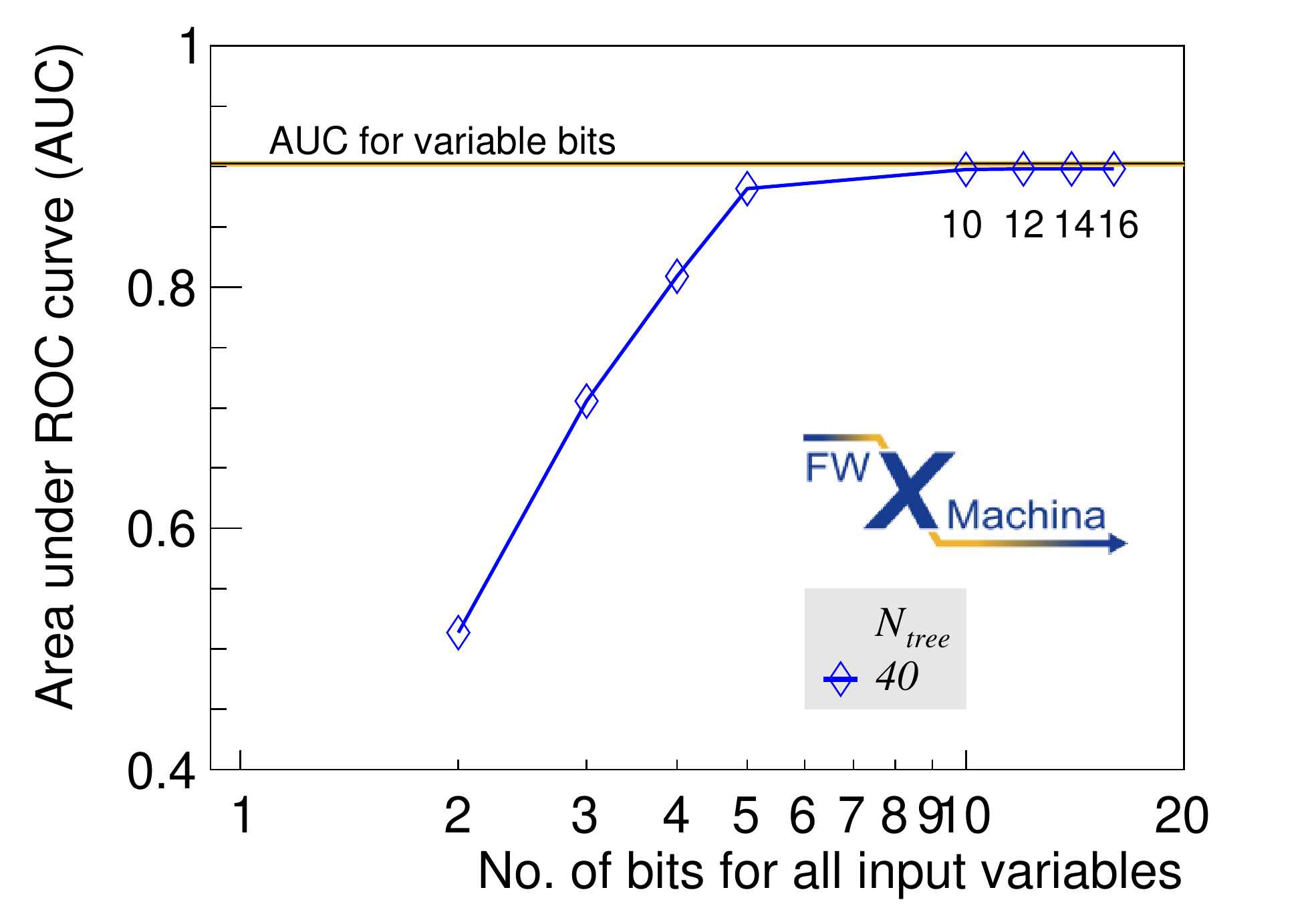}
    \caption{
    Physics performance vs.\ maximum tree depth $D$ (left) and vs.\ number of bits for input variables (right).
    The performance is given by area under the ROC curve$^{\ref{fn:auc}}$ (AUC) for each configuration.
    }
    \label{fig:auc_vs_d}
\end{figure}

We now discuss the trade-off between physics and engineering performance.
As can be seen in figure \ref{fig:auc_vs_d} the area under the ROC curve (AUC) is plotted vs.\ maximum tree depth $D$ and number of bits for input variables.\footnote{
  AUC is defined to be the area under the curve when plotting $\varepsilon_S$ vs.\ $\varepsilon_B$.
  \label{fn:auc}
}
An AUC of $0.5$ corresponds to the worst possible performance,
while an AUC of $1.0$ corresponds to a perfect discrimination between signal and background.
For the former,
the quick rise is followed by a plateau starting around $D=7$.
For the latter,
the plateau begins at around $5$.

\subsubsection*{FPGA cost and parameter scanning}

\begin{table}[p!]
\caption{\label{table:benchmark}
    Benchmark configuration and the FPGA cost.
    Three groups of information are given.
    The top-most group defines the FPGA setup.
    The second group defines the ML training used for the $\MET$ problem
    and the Nanosecond Optimization.
    The third group gives the actual results measured on the FPGA for four tree-depth combinations of 
    $40$-$5$,
    $40$-$6$,
    $20$-$7$,
    and
    $10$-$8$.
}
\centering
{\small
\begin{tabular}{
    p{0.32\textwidth}
    p{0.30\textwidth}
    p{0.30\textwidth}
    }
\hline
Parameter                                   & Value & Comments \\
\hline
FPGA setup \\
    \quad Chip family                       & \multicolumn{2}{l}{Xilinx Virtex Ultrascale+} \\
    \quad Chip model                        & \multicolumn{2}{l}{xcvu9p-flga2104-2L-e} \\
    \quad Vivado version                    & \multicolumn{2}{l}{2019.2} \\
    \quad Synthesis type                    & \multicolumn{2}{l}{C synthesis} \\
    \quad HLS or RTL                        & \multicolumn{1}{l}{HLS} &
    HLS interface pragma: None \\
    \quad Clock speed                       & $320\,\textrm{MHz}$     & Clock period is $3.125\,\textrm{ns}$ \\
\hline
\multicolumn{3}{l}{ML training configuration \& Nanosecond Optimization configuration} \\
    \quad ML training method                & Boosted decision tree   & Regression, Adaptive boosting \\
    \quad No.\ of input variables           & $8$                     & \\
    \quad \textsc{Bin Engine} type          & \multicolumn{2}{l}{\textsc{Deep Decision Tree Engine} (DDTE)} \\
    \quad No.\ of bits for all variables  & $16$ bits for each & binary integers \\
\hline
FPGA cost for 40 trees, 5 depth\\
    \quad Latency                   & $6$ clock ticks             & $18.75\,\textrm{ns}$  \\
    \quad Look up tables            & $1675$ out of $1\,182\,240$ & $0.1\%$  of available \\
    \quad Flip flops                & $1460$ out of $2\,364\,480$ & $<0.1\%$ of available \\
FPGA cost for 40 trees, 6 depth\\
    \quad Latency                   & $9$ clock ticks             & $28.125\,\textrm{ns}$ \\
    \quad Look up tables            & $4566$ out of $1\,182\,240$ & $0.4\%$ of available \\
    \quad Flip flops                & $2516$ out of $2\,364\,480$ & $0.1\%$ of available \\
FPGA cost for 20 trees, 7 depth\\
    \quad Latency                   & $15$ clock ticks            & $46.875\,\textrm{ns}$ \\
    \quad Look up tables            & $4568$ out of $1\,182\,240$ & $0.4\%$ of available \\
    \quad Flip flops                & $2697$ out of $2\,364\,480$ & $0.1\%$ of available \\
    \quad Block RAM                 & $4.5$ out of $4320$         & $0.1\%$ of available \\
FPGA cost for 10 trees, 8 depth\\
    \quad Latency                   & $21$ clock ticks            & $65.625\,\textrm{ns}$ \\
    \quad Look up tables            & $2556$ out of $1\,182\,240$ & $0.2\%$ of available \\
    \quad Flip flops                & $2299$ out of $2\,364\,480$ & $0.1\%$ of available \\
    \quad Block RAM                 & $5$ out of $4320$           & $0.1\%$ of available \\
\multicolumn{3}{l}{Common values for the above configurations} \\
    \quad Interval                  & $1$ clock tick              & $3.125\,\textrm{ns}$  \\
    \quad Block RAM                 & $0$ out of $4320$           & If not listed above \\
    \quad Ultra RAM                 & $0$  out of $960$           & Same for all trees and all depth\\
    \quad Digital signal processors & $0$  out of $6840$          & Same for all trees and all depth\\
\hline
\end{tabular}
}
\end{table}

FPGA cost denotes the timing results,
consisting of latency and interval,
and resource usage.

Starting from our benchmark configurations listed in table~\ref{table:benchmark},
BDT parameters are varied,
one at a time,
to investigate cost dependencies on tuneable parameters.
We focus on the maximum tree depth $D$,
and also show dependencies on the number of bits.
We note that the maximum BDT complexity scales with $2^D$,
but,
as discussed previously in section~\ref{sec:nano_opt},
figure~\ref{fig:bins_vs_d} showed that the scaling is much softer,
especially at high $D$ values.

\begin{figure}[hbtp!]
    \centering
    \includegraphics[width=0.495\textwidth]{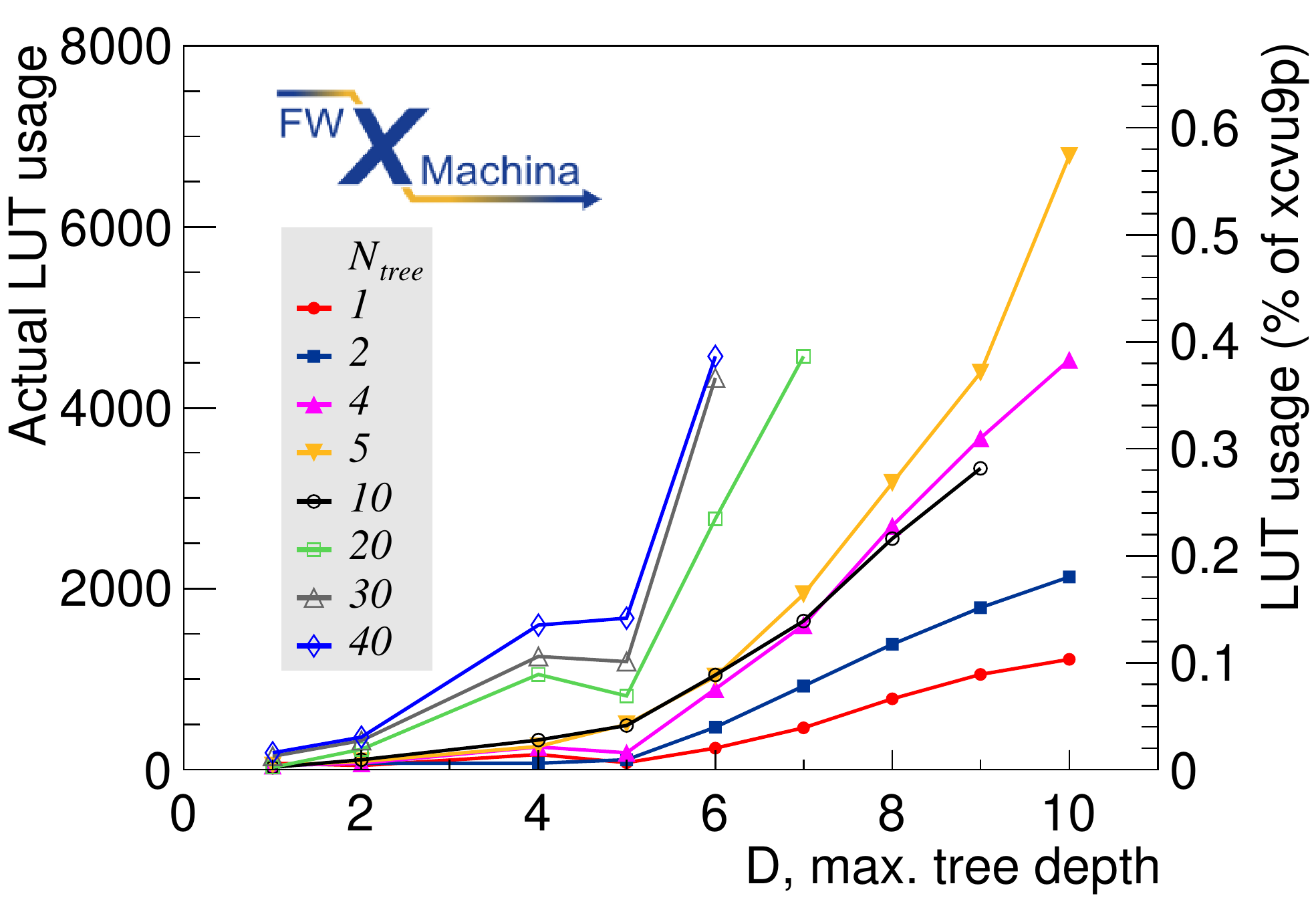}\hspace{0.01\textwidth}%
    \includegraphics[width=0.495\textwidth]{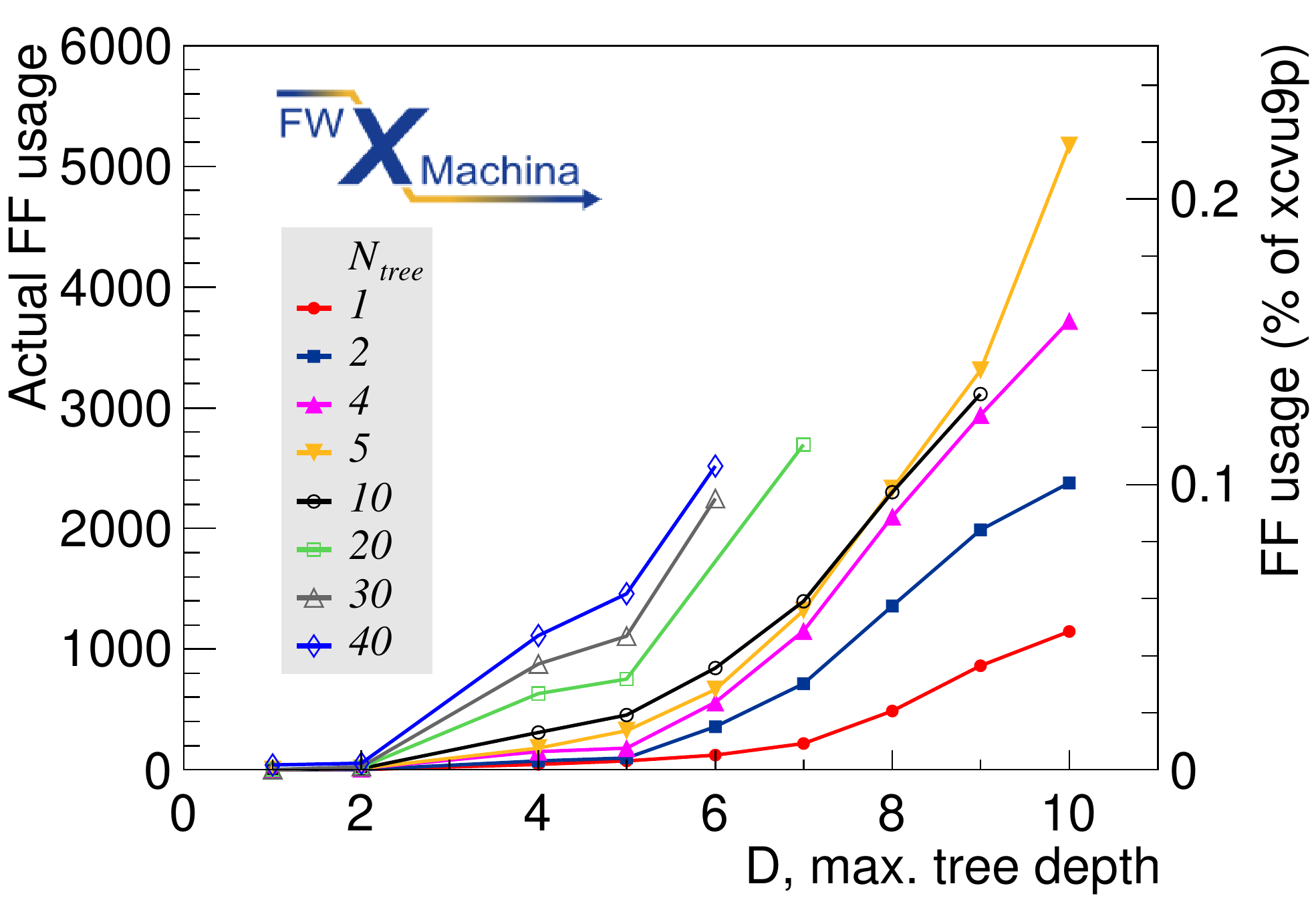}
    \caption{
    Actual LUT usage (left) and actual FF usage (right) as a function of the maximum depth.
    Absolute usage is shown on the left axis and percentage of our FPGA resources is shown on the right axis,
    both using the setup in table \ref{table:benchmark}.
    }
    \label{fig:lut_vs_depth}
\end{figure}

\begin{figure}[hbtp!]
    \centering
    \includegraphics[width=0.495\textwidth]{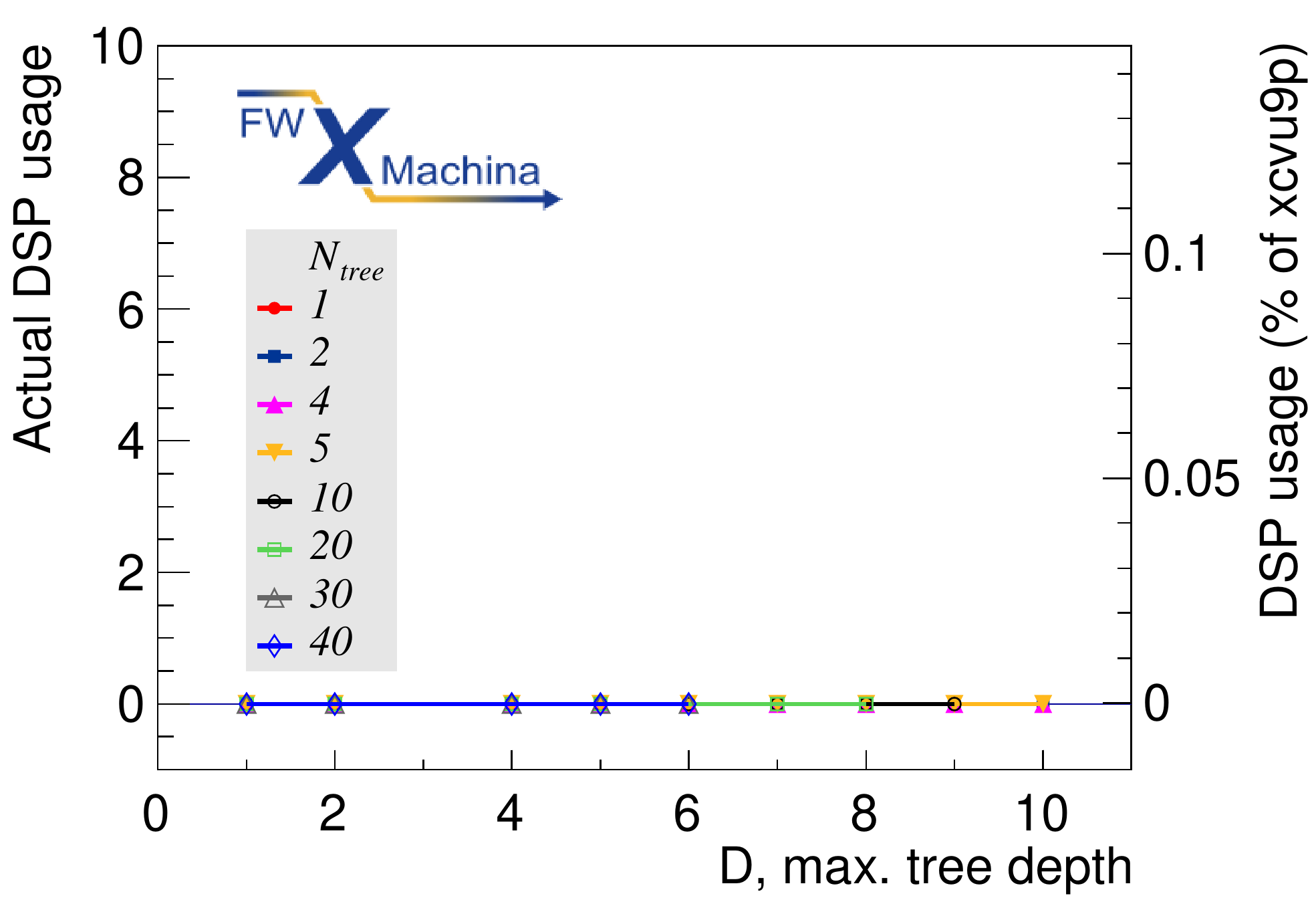}\hspace{0.01\textwidth}%
    \includegraphics[width=0.495\textwidth]{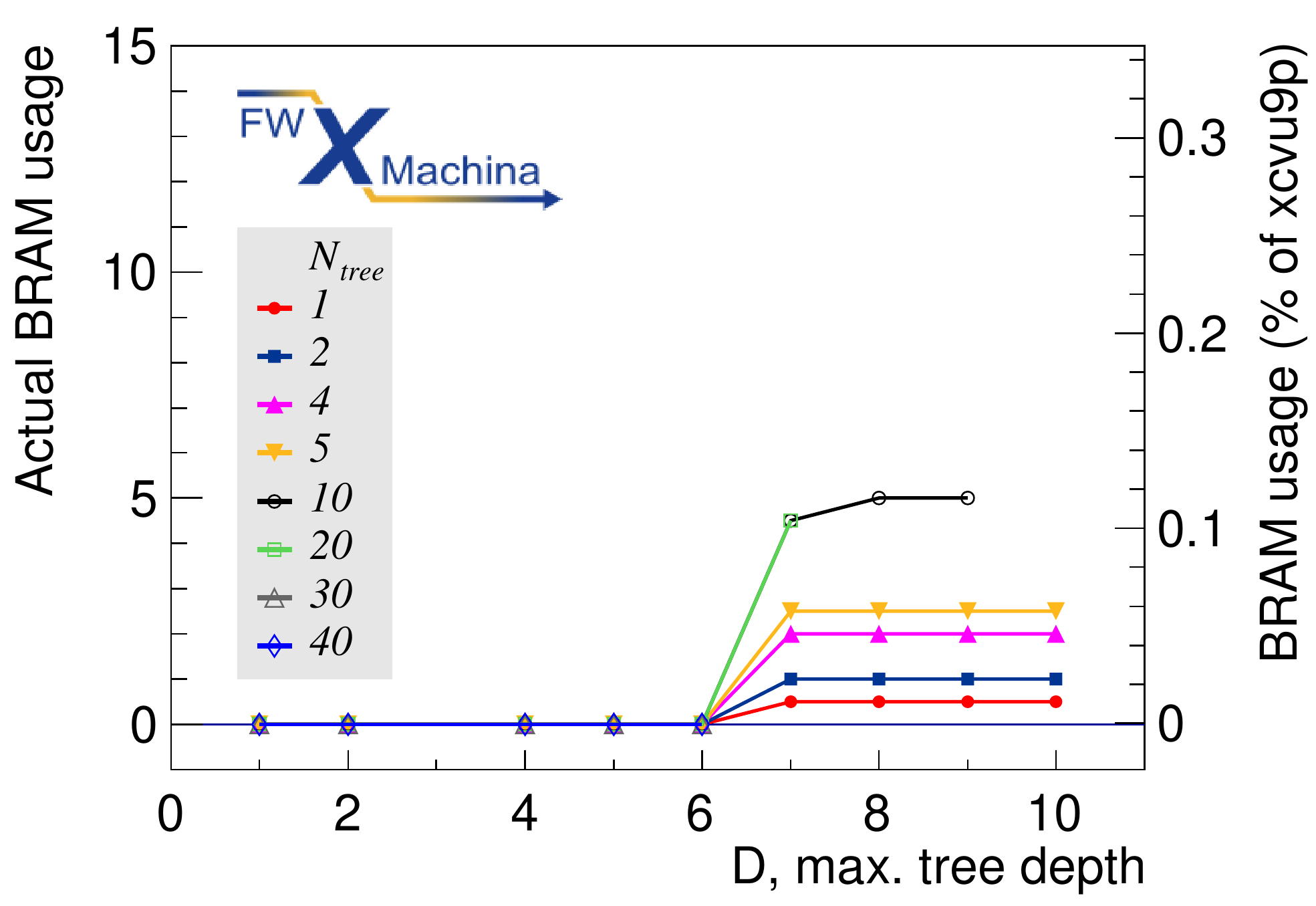}
    \caption{
    Actual DSP usage (left) and actual BRAM usage (right) as a function of the maximum depth.
    Absolute usage is shown on the left axis and percentage of our FPGA resources is shown on the right axis,
    both using the setup in table \ref{table:benchmark}.
    No DSP usage is seen.
    }
    \label{fig:dsp_vs_depth}
\end{figure}

The resource usage and latency scaling follows a similar pattern.
The number of look-up tables and flip-flops used vs.\ $D$ for several values of $N_\text{trees}$ is shown in figure~\ref{fig:lut_vs_depth}.
DSP usage is at $0$ for all configurations and BRAM is minimal as shown in
figure~\ref{fig:dsp_vs_depth}.

\begin{figure}[hbtp!]
    \centering
    \includegraphics[width=0.495\textwidth]{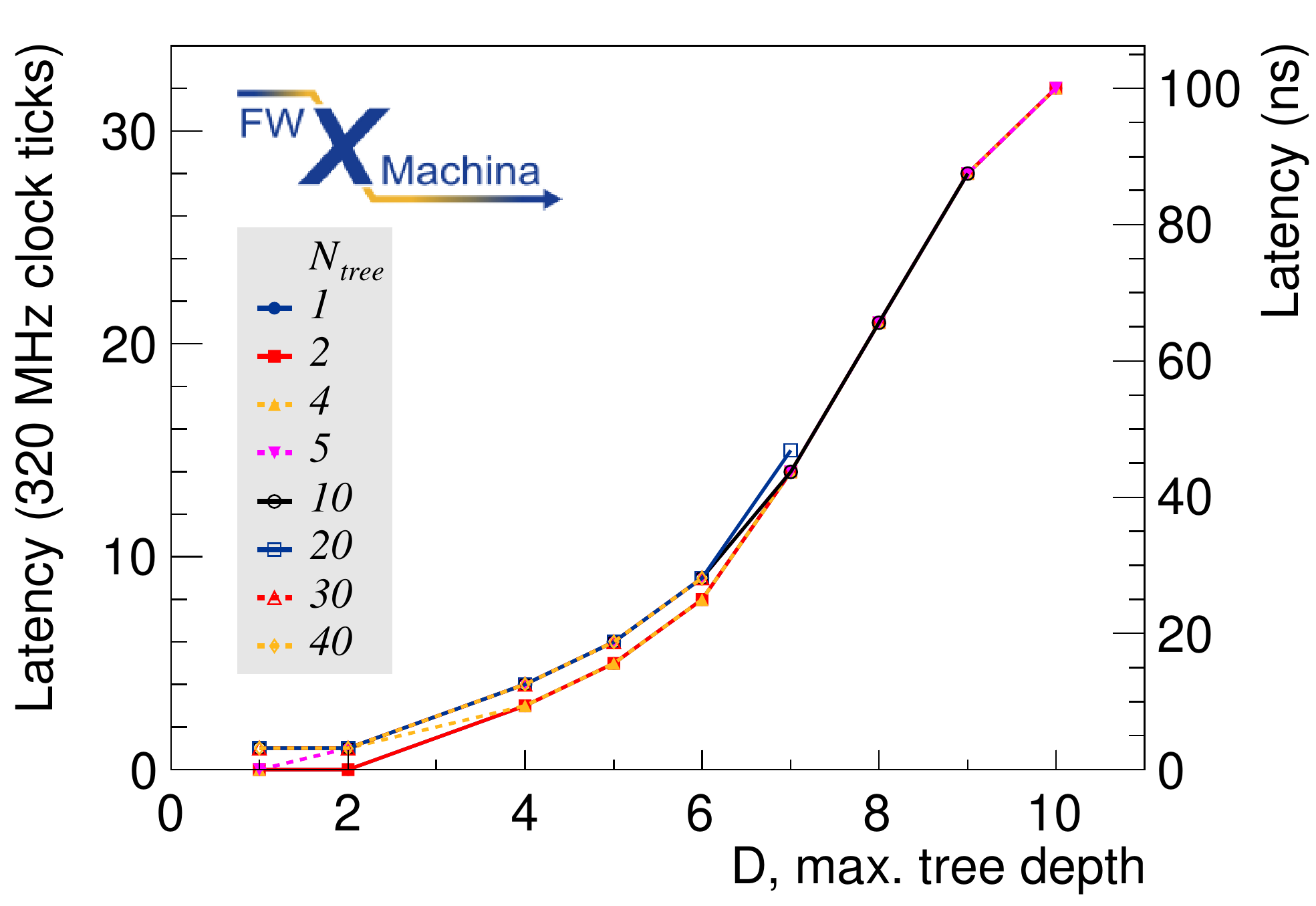}\hspace{0.01\textwidth}%
    \includegraphics[width=0.495\textwidth]{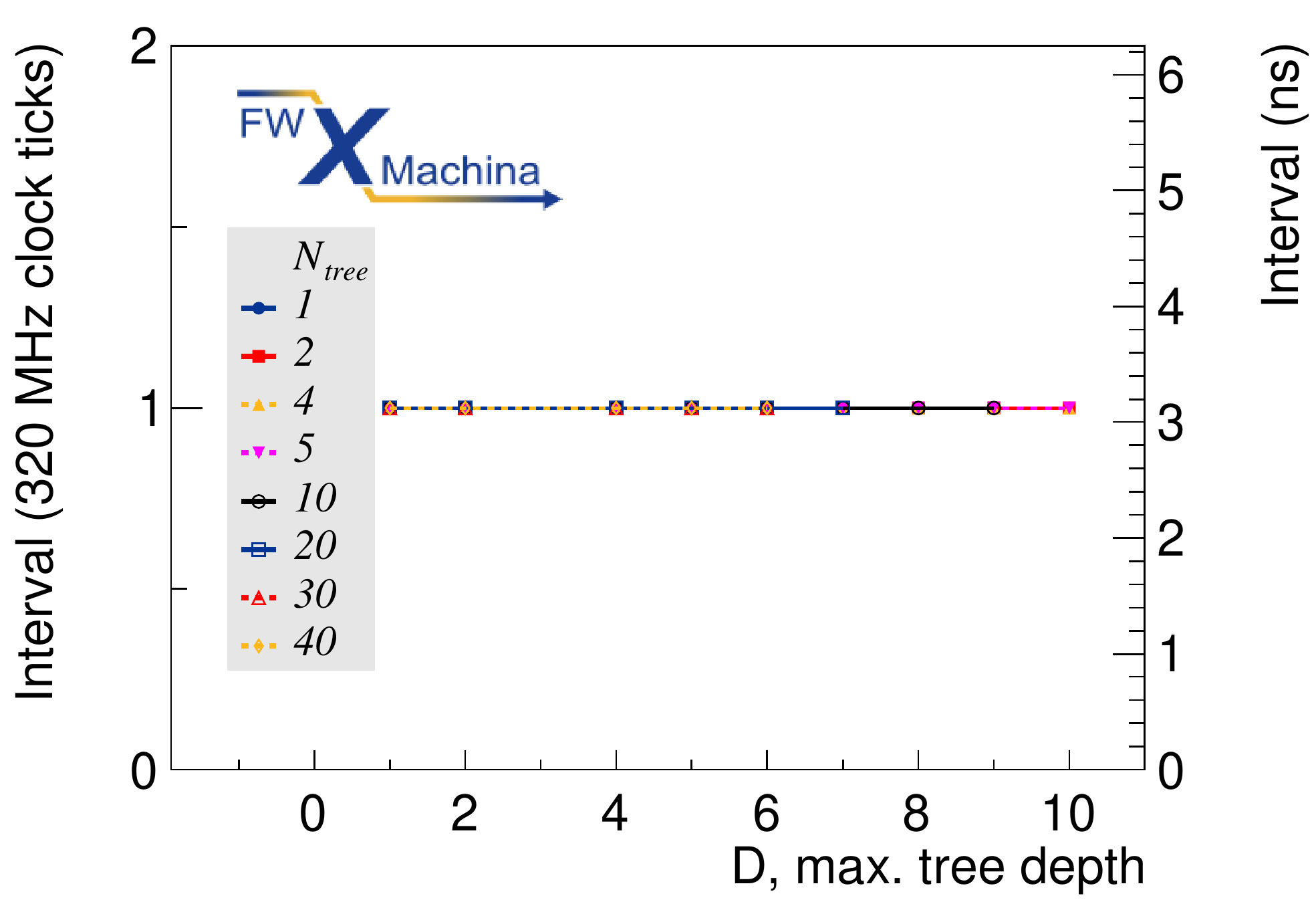}
    \caption{
    Algorithm latency (left) and interval (right) as a function of the maximum depth.
    Clock ticks are shown on the left axis and nanoseconds are shown on the right axis,
    both using $320\,\textrm{MHz}$ clock speed.
    Data series for the a given number of trees are connected.
    The interval is unity for all data points.
    Eight input variables of $16$-bit precision are used.
    }
    \label{fig:latency_vs_depth}
\end{figure}

The latency dependency on the maximum tree depth shows a similar pattern in figure~\ref{fig:latency_vs_depth},
Notably,
this figure also demonstrates that the number of trees does not seem to have a large significant impact on the latency,
with configurations from $1$ to $40$ trees all remaining within a single clock tick of each other at each maximum depth.
As in our previous firmware design,
the interval is only one clock tick for all configurations.

\begin{figure}[hbtp!]
    \centering
    \includegraphics[width=0.495\textwidth]{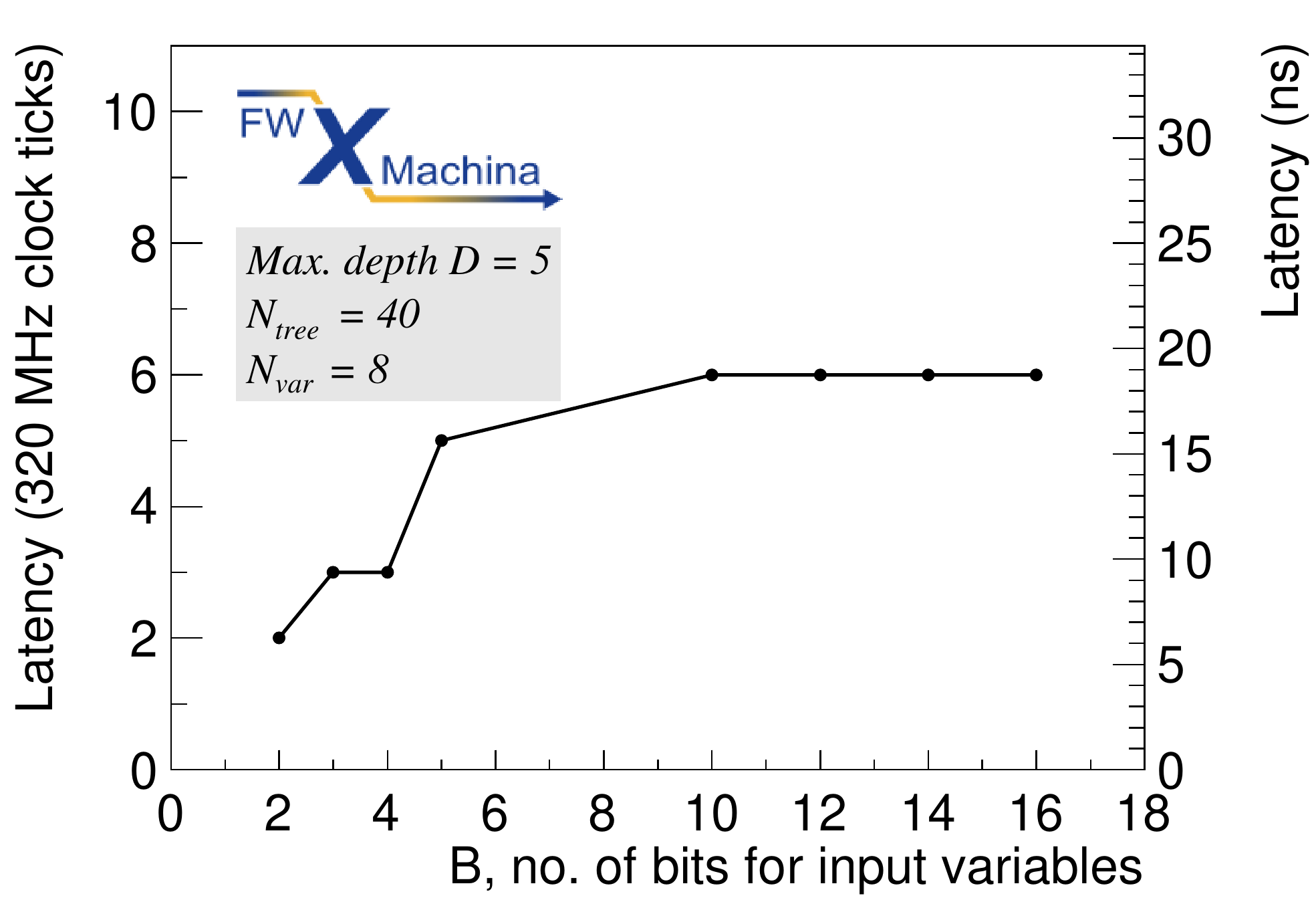}
    \caption{
    Algorithm latency as a function of the number of bits of the input variables.
    Clock ticks are shown on the left axis and nanoseconds are shown on the right axis,
    both using $320\,\textrm{MHz}$ clock speed.
    For the other parameters,
    Eight input variables of $40$ trees with a maximum depth of $5$ are used.
    }
    \label{fig:latency_vs_bits}
\end{figure}

The latency dependency on the number of bits used in the input variable representations is shown in figure~\ref{fig:latency_vs_bits}.
Less precise variable representations result in lower latency.
As is shown previously in ref.~\cite{Hong:2021snb},
less precise variable representations often result in degraded ML performance.
Appendix~\ref{app:bits} includes a more in-depth discussion of dependency on integer precision and this trade-off.

\section{Conclusions}
\label{sec:conclusion}
We present a novel implementation of boosted decision trees on FPGA within the \fwX\ framework \cite{Hong:2021snb} that allows for deep decision trees.
In this paper,
we demonstrate the use case for the deep tree structure for regression problems.
The new firmware design makes use of parallel decision paths (PDP) to allow for deeper trees as well as arbitrarily many variables:
two limitations of the flattened decision tree structure of ref.~\cite{Hong:2021snb}.
Finally,
support for varying bit integer precision per variable is implemented,
allowing for further resource usage optimization.

The performance is shown for the problem $\ETmiss$ estimation.
It is shown that combining several conventional $\MET$ calculations with a regression BDT provided a better signal efficiency and background rejection for reasonable operating point for level-0\,/\,level-1 trigger systems at the LHC.
FPGA implementation details are provided for hundreds of configurations,
We find that latency results are $\mathcal{O}(10)\,\textrm{ns}$.
The resource usage is $\mathcal{O}(0.1)\%$ of those available on our FPGA with the important exception that no DSP resources are used.
Results for various configurations by scanning the BDT parameters---%
such as the number of trees,
the maximum tree depth,
and the number of bits---%
show that our implementation can be adapted for a variety of use cases

\FloatBarrier 

\appendix

\section*{Acknowledgments}
We thank Stephen Racz for the initial effort of the project.
We thank Gracie Jane Gollinger for computing infrastructure support.
We thank Joerg Stelzer for the TMVA discussions.
We thank Pavel Serhiayenka and Kushal Parekh for their assistance with FPGA data collection.
We thank the University of Pittsburgh for the support of this project,
especially for Brandon Eubanks and Emre Ercikti from the Electronics Shop of the Shared Research Support Services of the Dietrich School of Arts and Sciences.
TMH was supported by the US Department of Energy [award no.\ DE-SC0007914].
STR was supported by the Emil Sanielevici Scholarship.
Patent pending.

\appendix

\section{Variable number of bits}
\label{app:bits}

One subtle difference between the classification and regression implementation is the effect of bit integers on output scores.
Due to the advantage of pre-computing and pre-normalizing bin values to their trees' boost-weights,
it is important that the conversion of floating point target variable outputs to integers respects addition,
i.e.,
$f(x_1 + x_2) = f(x_1) + f(x_2)$.
This requirement only applies to the target variable,
not the input ones.
This is discussed in ref.~\cite{Hong:2021snb}.

For our classification application in ref.~\cite{Hong:2021snb},
the mapping is relatively straight forward.
The purity values ranging from $[0,1]$ could be scaled to a $B$-bit integer by multiplying by $2^B -1$ to achieve a range of $[0, 2^B -1]$.
The same scaling was applied to Yes/No Leaf outputs of $[-1,1]$ to achieve the range of $[-(2^B -1),2^B -1]$.
Such simple scaling is closed under addition,
allowing for the summation of output scores from a forest of trees. 

However,
in regression we are not promised such neat outputs.
In principle,
the target variable can be bounded between any two values or even be unbounded.
Luckily in high energy physics,
physical variables are often either conveniently bounded below by zero or be symmetrical. 
Energy can range from $[0,E_\textrm{max}]$ to be scaled to $[0,2^B -1]$ as before.
Momentum can range $[-p_\textrm{max}, p_\textrm{max}]$ to be scaled to $[-(2^B -1),2^B -1]$.
If a given variable are not
bounded or symmetric,
we adjust the range accordingly;
see table~\ref{table:bit-conversion} for examples.

\begin{table}[htbp!]
\caption{
    \label{table:bit-conversion}
    Bit integer conversion methods used for the target training variable.
    Four representative examples are given.
    }
\centering
{\small
\begin{tabular}{
    p{0.200\textwidth}
    p{0.200\textwidth}
    p{0.200\textwidth}
    p{0.250\textwidth}
}
\hline
\makecell[l]{Adjustment method }
& \makecell[l]{Initial range }
& \makecell[l]{Adjusted range}
& \makecell[l]{Adjusted bit range} \\
\hline
Positive & $[0, 22] $     & $[0,22]$     & $[0, 2^B -1]$ \\
Positive & $[50, 450] $   & $[0,450]$    & $[0, 2^B -1]$ \\
Symmetric& $[-120, 70] $  & $[-120,120]$ & $[-(2^B-1), 2^B -1]$ \\
Symmetric& $[-50, 70] $   & $[-70,70]$   & $[-(2^B-1), 2^B -1]$ \\
\hline
\end{tabular}
}
\end{table}

In some cases these methods may necessitate a very high precision.
For instance,
if a variable ranges between $[1\,000\,000, 1\,000\,001]$,
after the conversion there will be many excess bit integers between $[0, 1\textrm{M}]$,
and so a very high precision will be necessary to capture the range of interest.
A similar issue will arise with asymmetric ranges such as from $[-0.5,6000]$.
While we claim that,
in most physics applications,
variables and their ranges are well defined so that such problems will not arise,
we recognize that this may not be generally true for every application.
Therefore,
in some cases clever unit manipulation or variable definition may be necessary.

\begin{table}[hbtp!]
\caption{Bit integer example for variables used at the LHC.
Two scenarios are considered.
The first set distributes 24 bits evenly among three variables.
The second set distributes 22 bits more optimally considering that the angular resolution at the first level is not $<0.1$ and that the energy resolution is higher.
\label{table:bit-integers}
}
\centering
{\small
\begin{tabular}{
    p{0.150\textwidth}
    p{0.200\textwidth}
    p{0.050\textwidth}
    p{0.175\textwidth}
    p{0.050\textwidth}
    p{0.175\textwidth}
}
    \hline
    &
    &\multicolumn{2}{l}{Evenly distributed}
    &\multicolumn{2}{l}{Optimally distributed} \\
    Variable
        & Range
        & Bits
        & Resolution
        & Bits
        & Resolution
        \\
    \hline
    $\pT$
        & $[-10,1023]\,\textrm{GeV}$
        & $\phantom{1}8$
        & $4\,\textrm{GeV}$
        & $12$
        & $250\,\textrm{MeV}$
        \\
    $\phi$ position
        & $[-3.14,3.14]$
        & $\phantom{1}8$
        & $\approx 0.025$
        & $\phantom{1}5$
        & $\approx 0.20$
        \\
    $\eta$ position
        & $[-4.9,4.9]$
        & $\phantom{1}8$
        & $\approx 0.04$
        & $\phantom{1}5$
        & $\approx 0.15$
        \\        
    \hline
\end{tabular}
}
\end{table}


\FloatBarrier 
\newpage
 
\addcontentsline{toc}{section}{References}
\bibliographystyle{JHEP}

\end{document}